\documentclass[prodmode,acmtcps]{acmsmall} % Aptara syntax

\usepackage[ruled]{algorithm2e}

\SetAlFnt{\small}
\SetAlCapFnt{\small}
\SetAlCapNameFnt{\small}
\SetAlCapHSkip{0pt}
\IncMargin{-\parindent}

\usepackage{flushend}
\usepackage{url}
\usepackage{pgfplotstable}
\usepackage{siunitx}
\usepackage{longtable}
\usepackage{tabu}
\usepackage{cleveref}

\usepackage{floatrow}
\newfloatcommand{capbtabbox}{table}[][\FBwidth]
\usepackage{url}
\usepackage[utf8]{inputenc}
\usepackage{amssymb}
\usepackage{amsmath}
\usepackage{subfigure}
\usepackage{eurosym}
\usepackage{cmap}
\usepackage[switch]{lineno}

\acmYear{2017} 
\acmVolume{1} 
\acmNumber{1} 
\acmArticle{1} 
\acmMonth{1} 
\acmPrice{\$15.00}
\doi{10.1145/3140235}

\begin{document}
\setcopyright{acmcopyright}
%\acmJournal{TCPS}

%\linenumbers
% Page heads
\markboth{M. Schmidt, et al.}{Cyber-Physical System For Energy Efficient Stadium Operation}

% Title portion
%\title{Energy Efficiency Gains in Stadium Operation by Context-Aware Supervisory Predictive Control: A Cyber-Physical System Approach}
\title{Cyber-Physical System For Energy Efficient Stadium Operation: Methodology And Experimental Validation}
\author{Mischa Schmidt
\affil{NEC Laboratories Europe, Lule\aa~University of Technology}
Anett Sch\"ulke
\affil{NEC Laboratories Europe}
Alberto Venturi
\affil{NEC Laboratories Europe}
Roman Kurpatov
\affil{NEC Laboratories Europe}
Enrique Blanco Henr\'{\i}quez
\affil{NEC Laboratories Europe}
}
% NOTE! Affiliations placed here should be for the institution where the
%       BULK of the research was done. If the author has gone to a new
%       institution, before publication, the (above) affiliation should NOT be changed.
%       The authors 'current' address may be given in the "Author's addresses:" block (below).
%       So for example, Mr. Abdelzaher, the bulk of the research was done at UIUC, and he is
%       currently affiliated with NASA.

\begin{abstract}
The environmental impacts of medium to large scale buildings receive substantial attention in research, industry, and media. %Our work studies the operational specifics of sports arenas. 
This paper studies the energy savings potential of a commercial soccer stadium during day-to-day operation. Buildings of this kind are characterized by special purpose system installations like grass heating systems and by event-driven usage patterns.
This work presents a methodology to holistically analyze the stadium's characteristics and integrate its existing instrumentation into a Cyber-Physical System, enabling to deploy different control strategies flexibly. In total, seven different strategies for controlling the studied stadium's grass heating system are developed and tested in operation. 
Experiments in winter season 2014/2015 validated the strategies' impacts within the real operational setup of the \emph{Commerzbank Arena}, Frankfurt, Germany. With 95\% confidence, these experiments saved up to 66\% of median daily weather-normalized energy consumption. Extrapolated to an average heating season, this corresponds to savings of 775 MWh and 148 t of CO$_{2}$ emissions. In winter 2015/2016 an additional predictive nighttime heating experiment targeted lower temperatures, which increased the savings to up to 85\%, equivalent to 1 GWh (197 t CO$_{2}$) in an average winter. Beyond achieving significant energy savings, the different control strategies also met the target temperature levels to the satisfaction of the stadium's operational staff.  
While the case study constitutes a significant part, the discussions dedicated to the transferability of this work to other stadiums and other building types show that the concepts and the approach are of general nature. Furthermore, this work demonstrates the first successful application of Deep Belief Networks to regress and predict the thermal evolution of building systems.
\end{abstract}





% We no longer use \terms command
%\terms{Design, Algorithms, Performance}


\acmformat{Mischa Schmidt, Anett Sch\"ulke, Alberto Venturi, Roman Kurpatov and Enrique Blanco Henr\'{\i}quez, 2016. Energy Efficiency Gains in Stadium Operation by Application-Aware Supervisory Control.}
% At a minimum you need to supply the author names, year and a title.
% IMPORTANT:
% Full first names whenever they are known, surname last, followed by a period.
% In the case of two authors, 'and' is placed between them.
% In the case of three or more authors, the serial comma is used, that is, all author names
% except the last one but including the penultimate author's name are followed by a comma,
% and then 'and' is placed before the final author's name.
% If only first and middle initials are known, then each initial
% is followed by a period and they are separated by a space.
% The remaining information (journal title, volume, article number, date, etc.) is 'auto-generated'.

\begin{bottomstuff}
The authors would like to thank Bilfinger HSG Facility Management and Stadion Frankfurt Management for technical support, consultation, and providing access to the data of the Commerzbank Arena, Frankfurt. The presented research work was funded by the European Commission within the \textit{Seventh Framework Programme} FP7 (FP7-ICT) as part of the \textit{Control \& Automation Management of Buildings \& Public Spaces in the 21st Century} (CAMPUS21) project under grant agreement 285729. 
Author's addresses: M. Schmidt, A. Sch\"ulke, A. Venturi, R. Kurpatov, {and} E. B. Henr\'{\i}quez, Smart Energy Research Group, NEC Laboratories Europe, 69115 Heidelberg, Germany.
\end{bottomstuff}

\maketitle

\section{Introduction}
Throughout the world, buildings are major consumers of energy producing significant amounts of Green House Gas emissions.  According to \cite{BuildingsEnergyBook2011}, residential and commercial buildings jointly accounted for 41\% of the US' primary energy use in 2010. Fossil fuels served close to 75\% of this consumption, with space heating (37\%), water heating (12\%), space cooling (10\%), and lighting (9\%) jointly accounting for more than two-thirds of the building consumption. In conventional buildings, irrespective of the type of construction, up to 90\% of energy is used during their operational phase \cite{Chau2015395}. %For low energy buildings, operation consumes up to 50\% of lifetime consumption. 
There are two complementary approaches to address the lion's share of building lifetime energy consumption: (i) refurbishments with better materials, components, and systems and (ii) improving operational strategies. Buildings already equipped with some level of automation infrastructure are particularly suitable for the latter approach by adopting a Cyber-Physical Systems (CPS) approach. In this approach ``computers and networks monitor and control the physical processes, usually with feedback loops where physical processes affect computations and vice versa'' \cite{Lee:EECS-2008-8}.

This work studies the energy savings potential of intelligent control strategies when applied to a commercial soccer stadium. Concretely, this work shows how the CPS concept can improve the operation of a special-purpose stadium system: the grass heating system of a German Bundesliga soccer stadium, the \emph{Commerzbank Arena} in Frankfurt am Main. This stadium was completely rebuilt for the FIFA World Cup 2006 \textsuperscript{TM}. Its age, installations, and capacity ($\approx$50,000 spectators) make it a typical representative of German stadiums. The stadium's grass heating system keeps the soccer pitch in high-quality condition throughout the winter season. Systems of this type account for a considerable share of the stadium's total thermal energy demand, but current natural gas prices in the range of \EUR{0.05}/kWh make them commercially viable. The alternative to heating the soccer pitch is to replace it during winter due to wear and tear with associated costs of \EUR{100,000} \cite{sportsturf}. 

Based on the holistic analysis of data obtained from the Commerzbank Arena’s automation system, we build a closed loop CPS controlling the grass heating system. This work documents the outcome of a series of experiments executed in two consecutive winters to validate the methodology applied in daily operation. The automated control strategies realize substantial savings in energy, associated cost, and CO$_2$ emissions while meeting requirements for grass growth to the staff's satisfaction. While the case study demonstrates the methodology's viability and the strategies' effectiveness, we also discuss the transferability of this work to other stadiums and other buildings. 

The remainder of the paper is organized as follows: Section \ref{sec:problem} introduces the problem statement, formulates the hypotheses underlying this work, and discusses related work. Section \ref{sec:methodology} presents the methodology and the concrete methods applied. Section \ref{sec:setting} introduces the Commerzbank Arena's heating system, derives the requirements specific to grass heating, and describes the communication platform deployed to interact with the stadium. Section \ref{sec:cps_rep} details the analysis of operational heating system data that makes predictive control possible. Section \ref{sec:intelligence} introduces seven heating strategies of various levels of complexity. Section \ref{sec:discussion} quantifies the experiments' impacts based on collected data. It also discusses the results and the potential limitations of this study, as well as the transferability to other stadiums and buildings in general. The paper concludes with a summary and an outlook on future work in Section \ref{sec:conclusion}.
% Head 1
\section{Problem Statement and Hypotheses, Related Work and Contribution, Terminology}
\label{sec:problem}
%-----------------------------------------------------------------
\subsection{Problem Statement and Hypotheses}
\label{sec:problemstatement}
A modern sports stadium has high energy requirements to provision systems like flood lighting, interior lighting, and catering, as well as supplying the event-specific media centers. Heating, ventilation, and air conditioning (HVAC) systems serve lounges and meeting spaces. Specific to lawn sports such as soccer, professional staff waters, lights, and cuts the grass throughout the year to meet the high pitch quality requirements - the stadium's prime target. For this purpose, the staff has access to information from sensors, the building systems, and services such as weather forecasts. In colder regions, it is common to heat the soccer pitch to maintain the grass at growth conditions throughout winter. Grass heating offers business value as it helps to avoid (i) pitch replacement during the season and (ii) costly match cancellations due to low pitch quality. As the heating also reduces the risk of player injuries, the German soccer regulations mandate the use of grass heating systems in the two top leagues \cite{DFBLO}. %In Austria, new regulations mandate grass heating for the top soccer league since season 2016/2017 \cite{ATLO} and in the US \cite{AmFoot}, several arenas are known to have under-soil heating systems installed. 
There are installations in other countries, e.g.~France, Russia, the UK, and the US in soccer, rugby, and American football stadiums.

During the heating season (October - March), the Commerzbank Arena's total heat demand can exceed the capacity of the main heating supply (a gas boiler-based hydronic system). As the grass heating system consumes up to 50\% of the peak output of the supply, it is the chief cause of heating shortages. When these occur, the stadium's heating distribution circuit serves the grass heating system with priority and sacrifices the service quality and thermal comfort of other areas like offices, the conference center, and meeting rooms. The current best practice to avoid shortages is to run the grass heating system only during nighttime and rely on the soil’s thermal inertia during the day. However, on cold days that does not suffice and the stadium has to resort to daytime grass heating to ensure grass quality, causing the described shortages. 

%In the , the stadium's overall heating . 

The objectives of this work are to improve energy efficiency, to maintain grass quality, and to mitigate heating shortages. %The present operational scheme is centered on human operators manually adjusting supply temperature set-point and system operation schedules. 
In the status quo operation, staff monitors the soccer pitch's quality visually and by using spot checks of the pitch's soil temperatures. Occasionally, staff adjusts heating system parameters manually. The frequency of checking varies with the workload, but usually, there are daily checks. In times of leave or of high workload, system operation is set to conservative settings to ensure grass growth even in the case of adversely changing weather conditions. This work explores the automation of adopting the staff's supervisory control decisions through reactive, predictive, and context-aware strategies by developing data-driven CPS capabilities, that leverage the stadium's Building Management System (BMS). %Predictive strategies require the creation of a computational representation of the physical process to be predicted. Conceptually, two kinds of approaches exist to do so: the formulation of models, e.g. based on physical properties, and data-driven approaches, e.g.~methods of the computational intelligence field. 

In summary, this work has two hypotheses:
\begin{enumerate}
\item \emph{The automation of currently manual supervisory control decisions improves efficiency in daily operation as less conservative operational settings are needed.}
\item \emph{Predictive and context-aware control strategies can mitigate heating shortages and further improve the building's operational efficiency.}
\end{enumerate}

\subsection{Related Work,  Relation to Earlier Work}\label{sec:related_work}
In theory, the grass heating system and the associated soil can be modeled based on physical parameters. However, \cite{Holmes-2008} argues that the standard soil temperature models \cite{Vries-1963,Cuaraglia-2001,Elias-2004} are not sufficiently precise to estimate the soil temperature near the surface. For this reason, the meteorology and geoscience communities provide several works on data-driven approaches for soil temperature prediction at different depths. Unfortunately, their focus lies on meteorological influences without discussing the possibility of under-soil heating systems.  For example, \cite{MET:MET1489} concentrates on day-ahead mean temperatures, whereas \cite{soilPredTurkey} and \cite{geographicalSoil} predict monthly mean temperatures. These works show that the daily mean air temperature is the dominant meteorological parameter impacting the soil temperature - solar radiation, relative humidity, wind speed, and precipitation play minor roles. Several studies focus on estimating soil temperatures at different depths by using either nearby weather stations' soil temperatures \cite{NeighborStationBased} or local meteorological data \cite{Tabari_ANN_MLR,DailySoil_Data_Spatial,Kisi_Soil_DiffDepths}. %Of the cited works, all but \cite{DailyNeuroFuzzy} study the application of neural networks.
All of the cited works study neural networks for predicting soil temperatures. 
In \cite{NeighborStationBased,Tabari_ANN_MLR,MET:MET1489} neural networks achieve higher regression accuracies when compared to linear or non-linear regression. Further, \cite{DailySoil_Data_Spatial} shows that neural networks outperform an adaptive inference system-based regression approach. The cited works focus on forecasting soil temperatures on timescales of one day or larger. However, to predictively operate the grass heating system, an intra-day prediction horizon is crucial. 

Stadium operation strongly depends on scheduled events, making it distinct from other medium or large-scale buildings studied in the building energy efficiency literature. Also, to the best of our knowledge, the recent research addresses neither the modeling nor the optimization of under-soil heating systems. 
However, the literature provides guidance on the methods and techniques to apply, as well as magnitudes of effect sizes to expect. As outlined in the following, current research on predictive building control strategies achieves high increases of performance by relying on predictive models learned from sensor data. 
 
\cite{Wei2015294} optimizes the operation of a multi-zone Heating, Ventilation and Air Conditioning (HVAC) system for room temperature and energy consumption, taking relative humidity, room temperature and indoor CO$_{2}$ levels as the input.  Compared to seven other regression models, a neural network ensemble performed best. A modified Particle Swarm Optimization algorithm solves for Pareto-optimal solutions of indoor air quality, comfort, and energy consumption by controlling the supply air's static pressure set-points. Different weightings of these objectives lead to different Pareto-optimal trade-offs. Regression models created from a recorded two week period indicate average estimated electricity savings of 12-17\%.

\cite{Ruano2016145} uses neural networks and multi-objective optimization for HVAC operation to minimize economic cost while ensuring user comfort. The study takes into account indoor temperatures, schedule information, cost, and weather variables. It documents energy consumption for three out of a total of six experiments conducted in winter and summer seasons at University of Algarve, Portugal. The experiments' lengths are relatively short with a maximum of two days. The results suggest financial savings while spending more energy to ensure minimized comfort violation: ``savings in the order of 50\% are to be expected''. 

Starting from a thermal building simulation, \cite{7317804} proposes to use neural networks to learn building behavior regarding energy and comfort subject to control actions. The Genetic Algorithm then derives building control rules. A knowledge base stores these, enabling facility managers e.g.~to strive for energy savings targets. The approach is verified in a care home in the Netherlands where heating supply, window opening, the degree of shading, and light levels can be controlled by using three months of simulation and two months of experimentation. Energy savings are normalized for weather influences using the Degree Day method and reach 25\%. 

\cite{Costanzo201681} uses an ensemble of neural networks to assist Reinforcement Learning in creating an HVAC demand response controller able to control on-off decisions. A simulation of 40 days with different temperature regimes validates the approach. %After collecting 16 data of days, the inferred control policies are stable within 90\% of the mathematical optimum. 
A shorter experiment in a living lab verifies the findings qualitatively.

The references indicate that neural networks are a popular regression technique in the meteorological, the geoscience, and the building optimization communities. The energy efficiency works show that validation is typically computational, or in case experimentation is used, it usually is limited to short periods of a few days or weeks. Furthermore, of the referenced works only \cite{7317804} uses weather normalization. None of the works uses methods of statistical inference. That, however, limits the generalizability and robustness of the results. The work in this paper relies on a prolonged observation and experimentation period spanning across three winters, accounts for weather influences on collected energy data, and applies methods of statistical inference to draw robust and reliable conclusions.

The lack of literature on intra-day soil temperature predictions subject to grass heating systems and the absence of accurate physical models advocate the application of a data-driven approach to the stadium's operational data. To extract the operational data needed and to also communicate control decisions to the grass heating system, this work creates a CPS leveraging the existing building instrumentation as much as possible.  % to derive the cyber-representation of the stadium soccer pitch. Doing so requires analysis of aggregated data captured from the stadium's Building Management System (BMS).
In larger facilities such as the Commerzbank Arena arena, staff typically relies on automation systems to operate building systems efficiently. 
Usually, the automation system architecture is three-layered \cite{Merz2009bas}: %as illustrated in Figure \ref{fig:bms}:
\begin{enumerate}
\item The lowest layer, the \emph{Field Layer}, consists of sensors and actuation devices.
\item The middle layer, the \emph{Automation Layer}, consists of controllers implementing control loops to meet configured set-points. 
\item The top layer, the \emph{Management Layer}, usually consists of the computer hosting the BMS. That allows monitoring building system operation and configuring set-points. Typically, these BMS provide basic means of configuration, e.g.~simple supervisory control rules and schedules.  
\end{enumerate}
We develop a CPS by extracting information from the Commerzbank Arena's automation infrastructure via the stadium's BMS and by accessing an internet weather forecast service. The information provides insights into the building operation and the associated physical processes of concern. That enables predictive or reactive control of building system operation parameters. The CPS issues appropriate control commands to the BMS to enact these using the lower automation infrastructure layers.

This paper builds on earlier findings in \cite{Schmidt:2015:EEP:2735960.2735978,Schmidt:2015:EEG:2821650.2821661} and extends these as follows:
\begin{itemize}
\item The present work describes the overall methodology that guided earlier work. Over the course of three years, its application formed an efficient and effective data-driven CPS integrating the Commerzbank Arena’s BMS and enabling flexible execution of different supervisory control algorithms.

\item \cite{Schmidt:2015:EEP:2735960.2735978} provides a description of the arena and its heating system. It provides the first analysis of data collected from the Commerzbank Arena's grass heating system captured in winter 2013/2014 by data aggregation platform deployed on top of the stadium's BMS. The analysis confirms literature in that air temperature is the primary meteorological parameter of interest impacting the soil temperature evolution in the absence of under-soil heating systems. The strong effect size allows reducing the number of the predictive models' input variables by neglecting other meteorological parameters than air temperature.  

Based on the collected operational data, the current work contributes the intervals of 95\% confidence for the grass heating system's energy consumption. Further, this work provides the accuracies of neural networks trained with the data to predict grass root temperature evolution in response to the heating system operation.  %These models support the predictive control strategies.

\item \cite{Schmidt:2015:EEG:2821650.2821661} formulates the seven different algorithms to control the stadium's grass heating system and provides their experimental validation - the winter 2014/2015 experiments. It provides descriptive statistics of weather-normalized energy use and grass root temperatures, demonstrating significant savings compared to the status quo operation. %In addition, an extrapolation of the energy savings to be expected in average winter conditions is provided. 

This work documents an additional experiment executed in winter 2015/2016 to quantify the effect of lowering the target soil temperature band. Further, for robust interpretation, it infers intervals of 95\% confidence of median normalized daily energy consumption for all Commerzbank Arena experiments and also infers the pairwise differences among the strategies' energy impacts. That allows the robust quantification of the differences in the strategies' effect sizes, i.e.~to reliably identify the most effective strategies.

\end{itemize}

\subsection{Notation}
Table \ref{tab:variables} summarizes the different variables used in this paper.  %Table \ref{tab:glossary} gives an explanation of terms.

\begin{table}[th!]
\centering
\tbl{{Variables used. The column \emph{BMS ID} specifies the BMS variable available to the CPS, if any.}\label{tab:variables}}{
\small
\begin{tabular}{r|l|l|l} 
Param. 	& Description & BMS ID &  Unit \\ \hline
$h$ & A specific dimension of $\vec{\hat{y}}$. Identifies a specific  & & $\{1,..~H\}$ \\
		& $T_{root}$ point estimate, $h$ time steps in the future & & \\
$H$ & Prediction horizon, depends on time resolution, & & $\in \mathbb{N}_+$ \\
		&  defines dimensionality of $\vec{\hat{y}}$ & &  \\
$HDD7$ 		& HDD for base temperature of 7$^\circ$C & & DD \\
HDI & Grass heating system \emph{Heat Demand Indicator}, &{H004-ST02} & 1/0 \\
		& indicates if system operates or not & & \\
$Q_{grass}$		& Measured energy use of grass heating system & {H004-ZM~UG1} & MWh \\
$\tilde{Q}_{grass}$		& Median of $Q_{grass}$ & & MWh \\
$Q_{grass,HDD7}$		& HDD7-normalized $Q_{grass}$ using Equation \ref{eq:final_hdd} & & MWh/DD \\
$\tilde{Q}_{grass,HDD7}$		&  Median of $Q_{grass,HDD7}$  & & MWh/DD \\
RMSE 				& Root Mean Squared Error & & K\\
$T_{external}$ 	&  Air temperature measured at time $t$& {H004-YB01} & $^\circ$C \\
$\overline{T_{external}}$ 		&  Daily mean air temperature &  & $^\circ$C \\
$T_{Gset}$ &  Grass heating supply temperature set-point & {H004-XS05} & $^\circ$C \\
$\Delta~T_{Gset,root}$ & $T_{Gset}-T_{root}$ & & K \\
$t_{now} $ & Current time & & Time\\
$t_{OFF}$ & Time of grass heating system deactivation & & Time \\
$t_{ON}$ & Time of grass heating system activation & & Time \\
$T_{root}$ &  Grass root temperature measured& {H004-ME20} & $^\circ$C \\
$\Delta~T_{root}$ &  Root temperature difference, e.g.~since $t_{ON}$ & & K\\
$\Delta~T_{root,ext}$ & $T_{root}-T_{external}$ & & K\\
$T_{supply}$ & {Main arena supply circuit temperature} & {WMZ01-ME3} & $^\circ$C \\
$\vec{\hat{y}}$ & Predictions of grass root temperatures in   & & $^\circ$C, $\in \mathbb{R}^H$ \\
				& defined forecast horizon & & \\	
\hline 
\end{tabular}
\normalsize
%\vspace{-0.5cm}
}\end{table}

% Head 2
\section{Methodological Approach}
\label{sec:methodology}

%-----------------------------------------------------------------
\subsection{Methodology}
This work develops a data-driven CPS to improve grass heating operation within its normal operational environment. The proposed approach makes sure to understand the requirements a control strategy needs to address and describes the process to build, deploy, and validate the CPS. %The approach also creates cyber-representation, and how control strategies are formulated and validated. 
The following steps form an understanding of the system encountered and the current best practice operation:
\begin{enumerate}
	\item \emph{Understanding the overall system, its operation, and data available.} Discussions with the arena's operational staff lead to a technical understanding of the grass heating system, its purpose, the thermal supply system, the ways of controlling operation, and the relevant data points available.\label{approach:step1} 
	\item \emph{Identification of requirements.}\label{approach:step2} By literature review and discussions with expert staff, a thorough understanding of the use case specific requirements is formed. 
	\item \emph{Establishment of communication for data extraction and actuation.} \label{approach:step3} 		
	This work pursues a data-driven approach relying on a data aggregation platform that supports the appropriate communication protocols to provide BMS access. That allows extracting building operation data and enacting actuation commands. As these influence the physical process, they impact future operational data and affect the future computational representation. Hence, this step creates a closed-loop CPS.
	%describes the communication platform that interfaces with the Commerzbank Arena's infrastructure to extract system operation data as well as to send control commands. 
	\item \emph{Data analysis and modeling of system characteristics.}  Monitoring the system in routine operation establishes a reference baseline and allows analyzing current control strategies to reveal inefficiencies. Moreover, the data allows modeling the soil characteristics for use in predictive control strategies.\label{approach:step4} %From April 2014 data, we are able to understand the soil's thermal characteristics without heating influence. 
	\item \emph{Development of improved control strategies.} Based on discussions with staff, the analyzed data, and the operational insights, control strategies are formulated.
	\label{approach:step5}
	\item \emph{Validation by experimentation.} 
	These experiments execute the different control strategies via the deployed platform within the real operational environment for prolonged periods of time. Data is recorded, analyzed, and discussed to extrapolate and generalize the results.\label{approach:step6} 	
\end{enumerate}
\subsection{Methods for Analysis and Inference}
This subsection provides information about the methods to be applied in the presented methodology steps \ref{approach:step4} and \ref{approach:step6}. Sections \ref{sec:cps_rep} and \ref{sec:discussion} document their application.

\subsubsection{Weather Normalization}
This work uses the Heating Degree Day (HDD) normalization technique %\cite{DegreeDays} 
to account for changing weather conditions across different years. It normalizes energy consumption $Q$ by dividing it by a normalization factor $HDD$ that captures the extent to which the measured outside air temperature $T_{external}$ is below a use case specific \emph{base temperature} $T_{HDD,base}$.  
This work follows the German standard \cite{VDI2067} by relying on daily mean air temperature ($\overline{T_{external}}$) for approximating HDD:
\begin{equation}\label{eq:final_hdd}
HDD\approx\begin{cases}
T_{HDD,base}-\overline{T_{external}} & T_{HDD,base}>\overline{T_{external}} \\
0 & T_{HDD,base}\leq~\overline{T_{external}}
\end{cases}
\end{equation} %The German standard HDD base temperature ($T_{HDD,base}$) for heating systems is defined as $15^\circ C$ \cite{VDI2067}. 
Usually, $T_{HDD,base}$ is defined as the outside air temperature below which the studied building requires heating. As the grass heating system is outdoors and 
%since the operational staff configured the system to become active only when the air temperature $T_{external}$ is below $7^\circ C$, 
the Commerzbank Arena's standard system configuration activates it only when $T_{external}\leq7^\circ C$, 
this work uses $T_{HDD,base}=7^\circ C$ for normalization. %The corresponding normalization factor is denoted as HDD7 and applied to the grass heating system's energy consumption $Q_{grass}$ to derive the normalized consumption $Q_{grass,HDD7}$. 
When calculating daily energy statistics, days with 0 HDD are excluded.% as $Q_{grass,HDD7}=\infty$.  

Degree-day-based calculations are especially sensitive to the choice of $T_{HDD,base}$ as it has a big effect on the proportional difference between different periods' HDDs (e.g.~days or winter seasons). Additionally, on days where $\overline{T_{external}}$ is close to the building's $T_{HDD,base}$, the building will often require little or no heating possibly leading to misleading or erroneous energy consumption statistics. We choose $T_{HDD,base}$ equal to the grass heating activation temperature. The stadium operations team, which is responsible for the arena's energy consumption, confirms this choice as appropriate. For robustness against multiplicative effects of HDD values close to 0, this work discusses energy-related effects using the median, not the mean, as described in the next section. 

Situations of intermittent heating, e.g.~around occupancy hours or as in the case of nighttime-only heating, are another aspect requiring consideration when applying HDD normalization. In these situations, the HDD value covering the full period (e.g.~a day) may not be a suitable representation of the air temperatures most relevant to the energy consumption. However, the thermal energy stored in and lost from the soccer pitch over the day is a result of $T_{external}$ and the soil's thermal inertia, i.e.~a multi-hour period. In particular, on days without daytime heating, i.e.~the days with mild temperatures on which best practice relies on nighttime-only heating, the grass heating system needs to counter daytime cool-down effects when the nighttime starts. %While the cooling curves in \cite{Schmidt:2015:EEP:2735960.2735978} exhibit a decline in speed after 6 hours, historic data of soil temperatures .
Thus, considering the full-day HDD is appropriate for describing the grass heating system's behavior even on days with nighttime-only heating.

\subsubsection{Descriptive Statistics and Statistical Inference}
For robustness against outliers, this work discusses effects on energy consumption and thermal behavior based on the \emph{median}. This measure of descriptive statistics describes the properties of the observed data but does not assume that the observations are samples of a larger population. However, this work interprets a change to control strategies as changing the underlying population's characteristics. Hence, to derive generalizable results from measured data, \emph{statistical inference for a single median} is used to characterize the data sets of the individual control strategies. This work uses statistical inference for \emph{two medians} to calculate a robust estimate of the pairwise differences between two strategies' effects while accounting for stochastic uncertainty. 

Specifically, this work applies inferential statistics using the Student's t-distribution to construct intervals of 95\% confidence in those cases that satisfy the underlying requirements of independence and normality of the collected data samples. While the former condition requires careful reasoning, the latter is tested using the Shapiro-Wilk test of normality. % provided in \cite{R}.
Where this test rejects the normality assumption, common statistical bootstrapping is used to construct the confidence intervals, using the percentile method with 10,000 bootstrap replicates. This work explicitly points out the situations of applying bootstrapping.  %\cite{boot}

\subsubsection{Assessing Grass Quality}\label{sec:grassquality}
After discussions with staff, the control problem is to keep the grass root temperature $T_{root}$ within defined temperature bands as much as possible. For this work, violations of minimum temperature levels are particularly critical as these affect the grass quality negatively. Thus, we assess the extent to which the control strategies fail to keep $T_{root}$ above a defined minimum temperature by relying on the service level Key Performance Indicator \emph{Under-Performance Ratio} (UPR). It represents the fraction of (operating) time the system does not meet minimum service level requirements. Other environmental parameters such as the soil's humidity are not monitored and thus unavailable for automated grass health monitoring. Throughout the three winters, spot checks by experts ensure that the pitch quality stays satisfactory. Section \ref{sec:discussion} incorporates the experts' feedback into its discussions. 

\subsection{Methods for Data-Driven Grass Root Temperature Prediction}
\label{sec:NN}

\cite{Schmidt:2015:EEP:2735960.2735978} identifies the dominant influencing variables for short term $T_{root}$ prediction. These are $T_{root}$ itself, $T_{external}$, information whether the heating system is being active ($HDI$), and the grass heating system's supply temperature (controlled by the corresponding set-point $T_{Gset}$ when the system is active). That study also characterizes the $T_{root}$ heating and cooling trends as non-linear, resembling a saturation curve.  
The current work applies the following two non-linear regression techniques, using fivefold cross-validation during model training.

\begin{itemize}	
	\item We focus on neural networks as \cite{Wei2015294} indicates that they outperform other non-linear standard regression techniques for thermal predictions. Section \ref{sec:related_work} shows that in particular, the feed-forward \emph{Multi-Layer Perceptron} (MLP) is popular when modeling thermal characteristics. 
	\item For comparison, this work also applies a \emph{Deep Belief Network} (DBN) \cite{Hinton:2006}. 
	As shown in \cite{salakhutdinov2015learning} and the references therein, DBNs have been applied successfully to visual object recognition, natural language processing, information retrieval, and robotics. This category of neural network consists of multiple stacked Restricted Boltzmann Machines that can be composed into deep network structures. During an unsupervised initialization phase, DBNs learn feature representations from inherent characteristics of the input data before learning to regress in a supervised way. To the best of our knowledge, these networks have not been applied to the field of predictive control of building systems, yet. 	
\end{itemize}

As not only a point estimate of the future is of interest, but also the temperature trajectory until that point, a vector $\vec{\hat{y}}$ of $T_{root}$ predictions is regressed. With a given time resolution, the length of prediction horizon $H$ defines the dimensionality of $\vec{\hat{y}}$. 

Both methods are implemented in Python with Theano \cite{Bastien-Theano-2012}. For both, the input data is standardized by shifting each feature datum by its training data mean and dividing by its training data standard deviation. 60\% of the data serves as the training set. The validation set and the validation set each use 20\% of the data. By using the same periods to train, validate, and test the regression models, weather parameters affect all models equally.  
A grid search identifies the different models’ hyper-parameter combination (e.g.~number of neurons, data history length, learning rate) performing best on the test set to tune the models to best performance. 
The standard Root Mean Squared Error (RMSE) metric assesses for each dimension of $\vec{\hat{y}}$ the regression model performance on the test set. $RMSE^h=\sqrt{\frac{1}{n}\sum_{i=1}^{n}{(y^h_i-\hat{y}^h_i)^2}}$, 
where $n$ is the number of predictions, $\hat{y}^h_i$ is the $i$\textsuperscript{th} value predicted by the model for the $h$\textsuperscript{th} time step into the future, and $y_i$ the $i$\textsuperscript{th} true value. %As our models predict the $T_{root}$ evolution of multiple intervals in a single prediction step, the term RMSE denotes a vector with dimensionality $H$. 

\section{Commerzbank Arena: The Thermal System, Requirements, and Communication Aspects}
\label{sec:setting}
\subsection{Methodology Step \ref{approach:step1}: System Understanding and Available Data}
Fig.~\ref{fig:system_design} depicts the Commerzbank Arena's hydronic heating distribution system. Two gas boilers with a total capacity of 2.4 MW supply the stadium with thermal energy. The main distribution circuit serves hot water to six different sub-systems. These are two domestic hot water systems (athletes’ showers and kitchen); two radiator-based static heating systems for east and west offices, lodges, and meeting rooms; the HVAC to supply warm air to all spaces with air conditioning; and the grass heating system. Waste heat is recovered from cooling machines and supplied to the hot water systems, primarily in summer. In the event of heating supply bottlenecks, the main circuit's temperature $T_{supply}$ drops below required levels.

\begin{narrowfig}{0.7\textwidth}[t]
\centering
\includegraphics[width=0.7\textwidth]{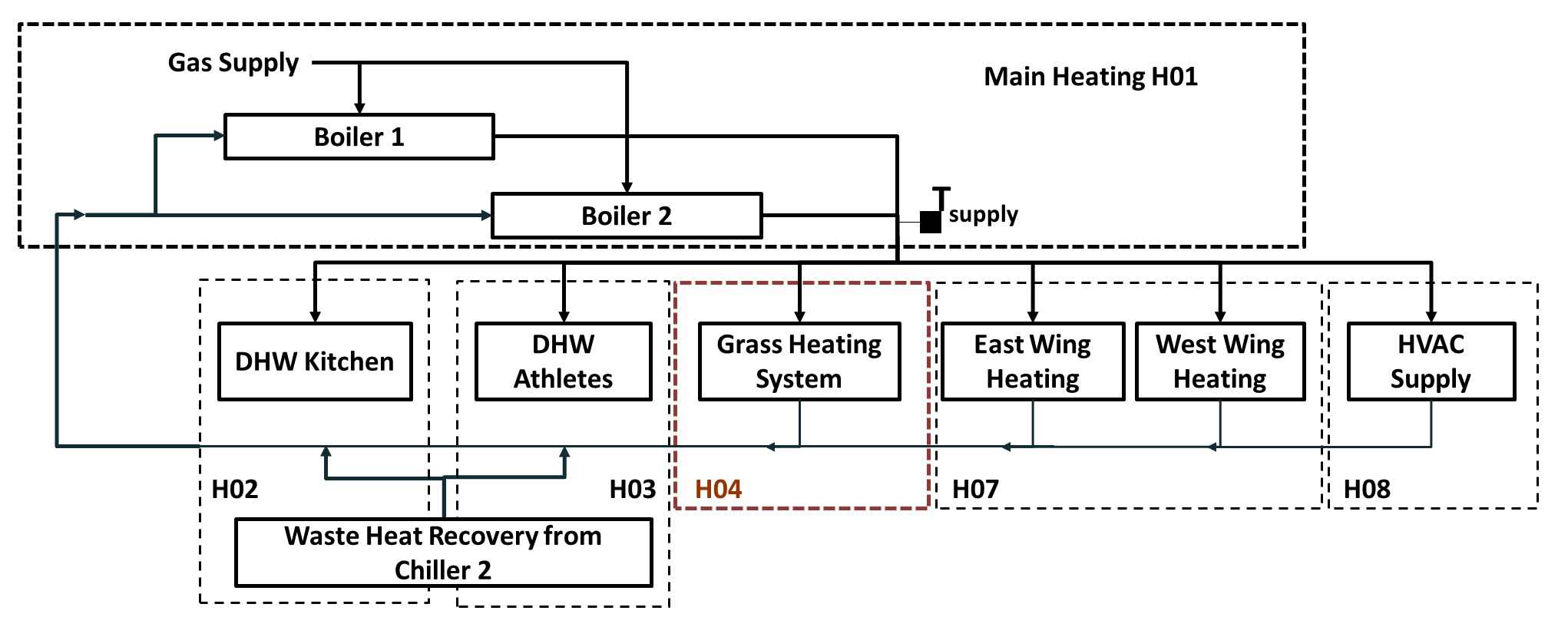}
\caption{The Commerzbank Arena's thermal distribution system, including the grass heating system H04.}\label{fig:system_design}
\end{narrowfig}

\begin{narrowfig}{0.5\textwidth}[t]
\centering
\includegraphics[width=0.5\textwidth]{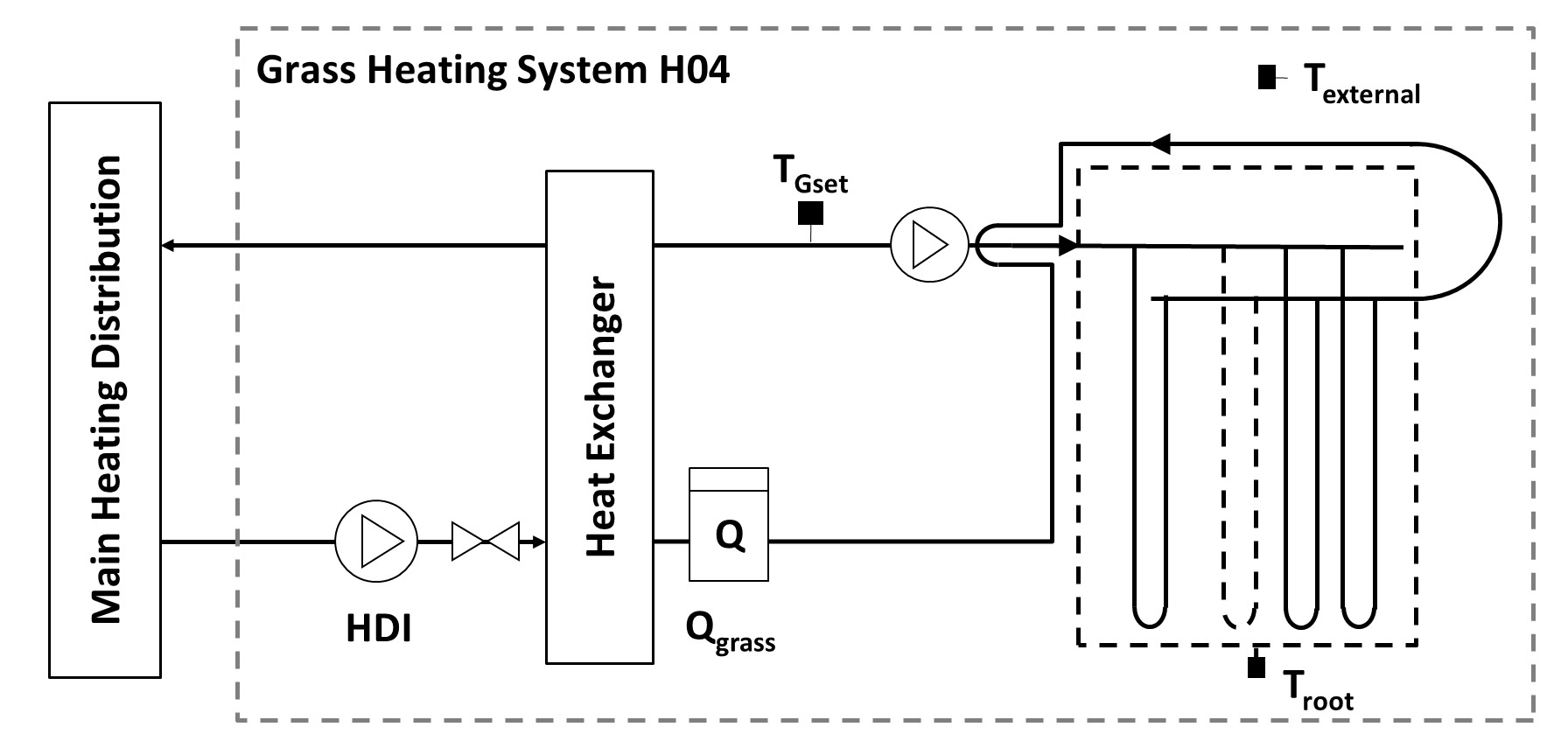}
\caption{Grass heating system H04 with sensor installations.}\label{fig:grass_sensor}
\end{narrowfig}

A 1.4 MW heat exchanger connects the stadium’s grass heating system shown in Fig.~\ref{fig:grass_sensor} to the main heating distribution network. 
Its multiple pipe loops distribute the water-glycol fluid under the playing field longitudinally. As indicated in Section \ref{sec:grassquality}, $T_{root}$ is the primary variable of concern to monitor the grass conditions. 
The corresponding sensor measures this variable in a depth of 15 cm. The BMS also monitors $T_{external}$, but it does not monitor other soil parameters such as humidity. When the heating system is active ($HDI=1$), $T_{Gset}$ controls the temperature of the fluid pumped through the pipe loops. That affects $T_{root}$ and consumes energy $Q_{grass}$. The system has a configurable fail-safe to protect the grass from overheating: the system deactivates if the fluid's temperature exceeds 40$^\circ C$ to avoid damaging the grass roots. Table \ref{tab:variables} indicates the stadium's relevant BMS variables.  
\cite{wunderground} provides weather forecasts for a nearby weather station (Frankfurt airport).

\subsection{Methodology Step \ref{approach:step2}: Grass Root Temperature Requirements}\label{sec:requirements}
To identify the requirements for grass heating, botanical literature and discussions with the arena's green keepers provide an understanding of the biological needs that drive the target $T_{root}$. In Germany, the Landscape Development and Landscaping Research Society e.V (Forschungsgesellschaft Landschaftsentwicklung Landschaftsbau e.V.) defines standard seed mixtures (``Regel-Saatgut-Mischungen'', RSM) for use in landscaping. German DIN 18035 \cite{DIN18035} recommends RSM categories 3.1 and 3.2 for sports use, consisting of a mixture of \emph{Lolium perenne} and \emph{Poa pratensis}. For these weeds, \cite{Beard2001} recommends $10^\circ C  \leq T_{root} \leq 18^\circ C$ for optimal growth. In coordination with the arena's experts, the $T_{root}$ target band for the winter 2014/2015 control experiments is defined as $12^\circ C$ to $14^\circ C$. This choice leaves a safety margin of 2K to the recommended minimum. For winter 2015/2016, the $T_{root}$ target is lowered to $10^\circ C$ to $12^\circ C$ to study the effects associated with removing the safety margin.

Note that in Germany, the different stadiums' staff meets several times per year to exchange experiences and best practices. Thus, we consider the Commerzbank Arena's best practice for $T_{root}$ regimes as representative. 

\subsection{Methodology Step \ref{approach:step3}: Establishment of Communication for Data Extraction and Actuation }
The data aggregation platform \cite{cesbpPaperC21} addresses Step \ref{approach:step3} of the presented methodology. The distributed and modular platform interacts with the Commerzbank Arena's BMS through the BACnet/IP protocol, enabling the development of data-driven CPS while maximally reusing the building automation infrastructure. The platform is designed to \emph{holistically} study the operational schemes and energy profiles of buildings by unifying data from different sources. A harmonized application interface serves the modules analyzing the data and implementing the control strategies. The BMS manages approximately 13,500 variables. These include readings from sensors and meters, as well as values of set-points, status flags, and internal values,  providing a detailed snapshot of the entire building state and its operation. Since August 2013, the platform accesses the BMS every 10 minutes. %This work focuses on the subset of BMS variables indicated in Table \ref{tab:variables}.
Table \ref{tab:variables} indicates the subset of BMS variables relevant for this work.

\section{Methodology Step \ref{approach:step4}: Data Analysis, Modeling of System Characteristics}
\label{sec:cps_rep}

\subsection{Current Control Schemes}
\begin{figure}[t]

\centering
\begin{tabular}{cc}

\subfigure[$T_{external}$, the dashed line indicates $T_{HDD,base}=7^\circ C$.\label{fig:winter1314_air_temps}]{
\includegraphics[width=0.45\textwidth]{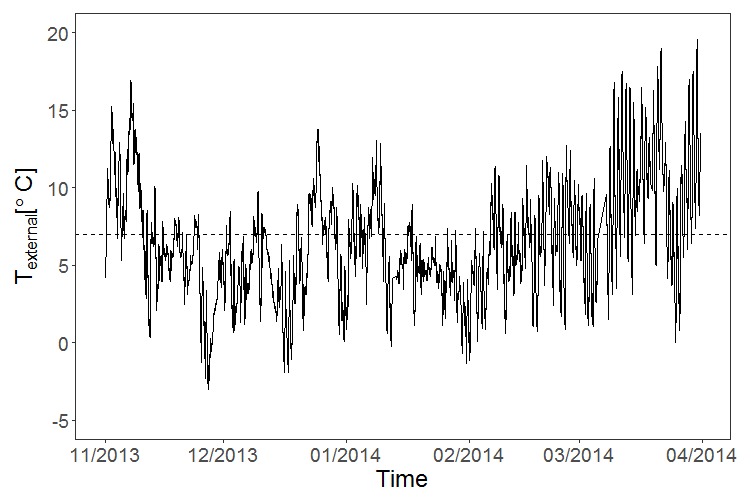}} & 

\subfigure[$T_{root}$ evolution, $T_{Gset}$ decided by stadium staff. Dashed lines indicate the first experiments' target band of $12^\circ C \leq T_{root} \leq 14^\circ C$.\label{fig:winter1314_grass_root_temps}]{
\includegraphics[width=0.45\textwidth]{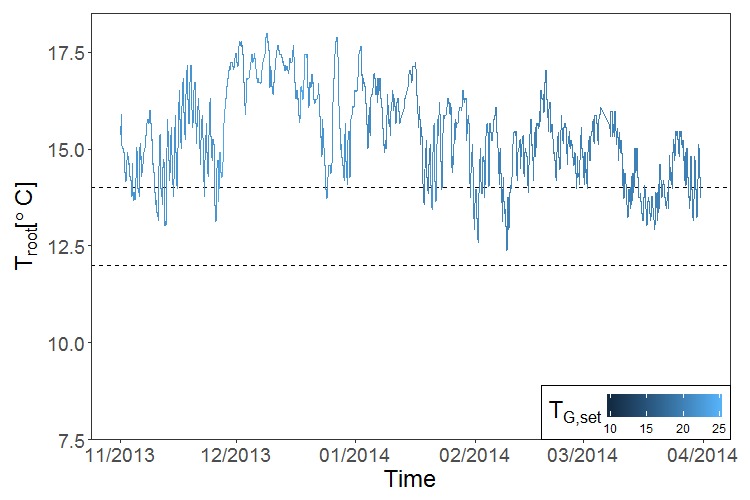}} 

%\caption{Grass root temperatures observed in reference winter 2013/2014.\label{fig:winter1314}}
\end{tabular}
\caption{Air and root temperatures observed in the stadium in reference winter 2013/2014.}\label{fig:winter1314_temps}
\end{figure}

\begin{figure}[t]
\centering
\begin{tabular}{cc}

\subfigure[Grass heating control in February 2014: temperature variables and grass heating operation indicator $HDI$.\label{fig:controlFeb14}]{
\includegraphics[width=0.65\textwidth]{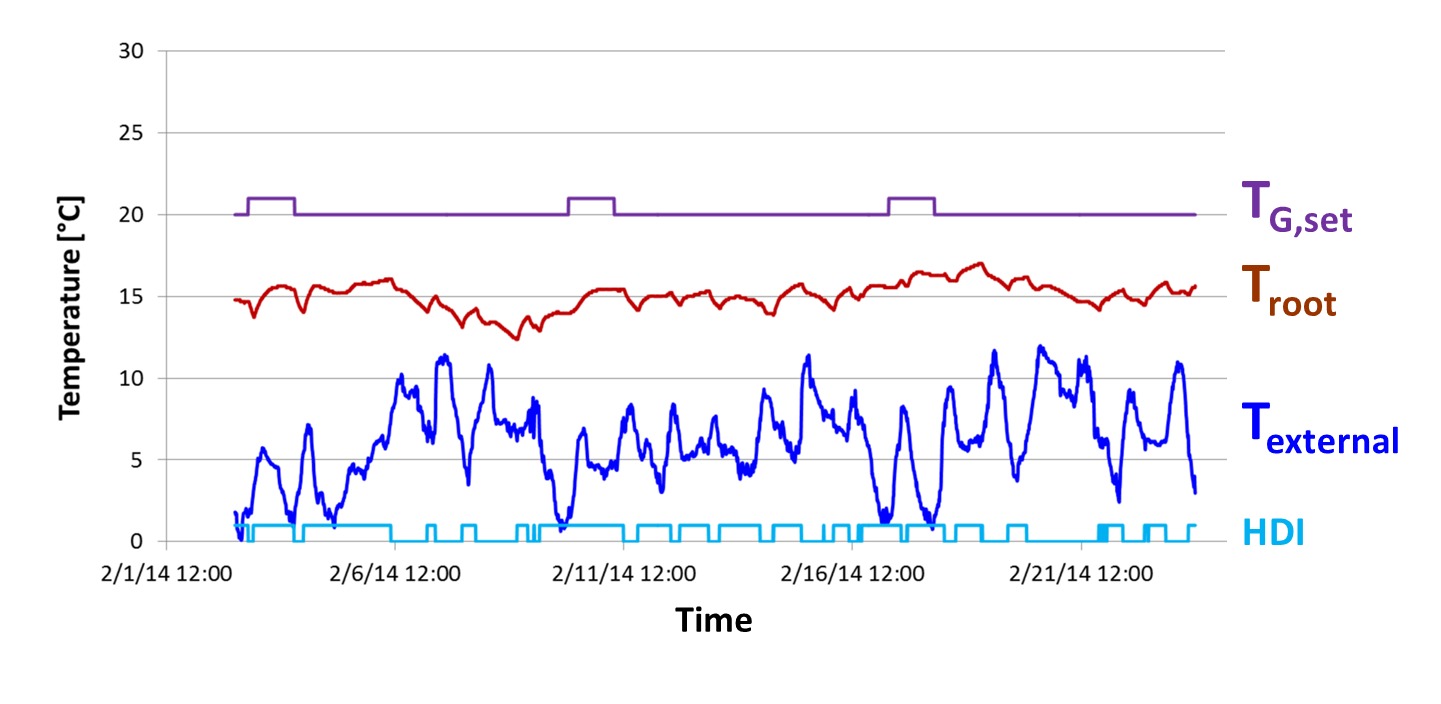}} &

\subfigure[$T_{root}$ distribution when grass heating active. Dashed lines indicate the target band of $12^\circ C \leq T_{root} \leq 14^\circ C$.\label{fig:root_temp_2013}\label{fig:refHistogram}]{
\includegraphics[width=0.3\textwidth]{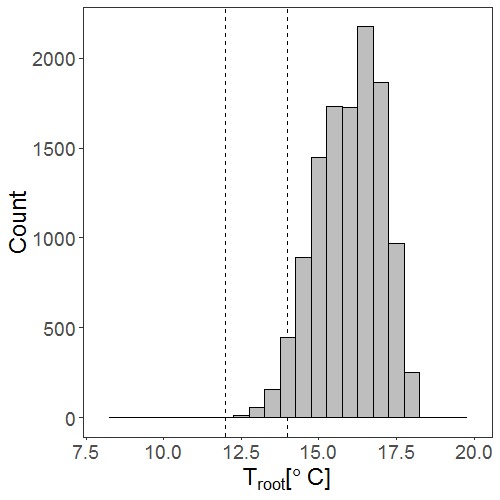}}

%\caption{Grass root temperatures observed in reference winter 2013/2014.\label{fig:winter1314}}
\end{tabular}
\caption{Status-quo operation resulted in wasteful heating in reference winter 2013/2014.\label{fig:winter1314}}
\end{figure}

In winter 2013/2014 the status quo grass heating operation consumed 795 MWh. Fig.~\ref{fig:winter1314_temps} depicts this reference period's recorded air and grass root temperatures. The status quo strategy is driven by manual operation as illustrated in Fig.~\ref{fig:controlFeb14}:
%The current grass heating operation is driven by human action as identified in \cite{ACMICCPS2015} and illustrated in fig.~\ref{fig:controlFeb14}: 
\begin{enumerate}
\item For days with mild temperatures, $T_{external}$ and time define the heating control. Heating is active only during nighttime (18:00 and 06:00) when $T_{external}\lessapprox 7^\circ C$. That operation schedule is best practice to avoid the reported shortages. 

\item Staff manually adjusts control parameters in several ways. For example, the data of February 2014 shows a regularly alternating $T_{Gset}$. On freezing days, staff changes from nighttime-only to daytime heating causing heating shortages in offices. 

\item Staff tends to choose conservative values for $T_{Gset}$, i.e.~higher values than necessary, as it cannot constantly monitor the system operation nor the grass conditions. 
\end{enumerate}

The observed control schemes' effects are visible in Fig.~\ref{fig:winter1314_grass_root_temps} showing higher than needed $T_{root}$.  Fig.~\ref{fig:root_temp_2013} focuses on $T_{root}$ for active grass heating only. Out of the total 1956 hours operating time, $T_{root}\geq 17^\circ C$ for more than 400 hours and $T_{root}\geq 15^\circ C$ for more than 1,600 hours. 
That presents an opportunity for significant savings.% by adapting the control parameters. %by modeling the thermal and energy behavior of the system for different modes of operation.

%\subsection{Observed Root Temperature Behavior and Energy Demand: General Observations}

%, reducing $T_{root}$ to the lower part of the recommended temperature ranges

\subsection{The Effects of Heating Activation} \label{sec:heating_activation}
\subsubsection{Statistical Considerations, Defining Heating Events}
The intention of this subsection is to understand the effects associated with an extended operation of the grass heating system. Its energy consumption, as well as its effect on $T_{root}$, are bigger after a prolonged cooling period than when the system had been heating shortly before. That establishes a time dependency among the data samples of $T_{root}$ prohibiting to apply statistical inference techniques to the samples directly. To establish independence among the samples and to capture prolonged heating effects, this work focuses on heating activation events at time $t_{ON}$ meeting the following criteria.

\begin{itemize}
\item The events start a period of active grass heating of 6 hours or longer. 
\item A period of inactive heating precedes the events, so that heat from the earlier heating cycle has been lost. That allows viewing the events as mutually independent.
\end{itemize}

Considering the second aspect, this work considers two alternatives of defining the period preceding $t_{ON}$. Both ensure heating event independence to study the effects on predictive accuracy. Definition \ref{def:unicool} is more restrictive than Definition \ref{def:intercool}. 
While Definition \ref{def:unicool} selects 87 heating system activation events during winter 2013/2014 (resulting in regression model training sets of 52 events),   Definition \ref{def:intercool} selects 117 events (leading to a training set size of 70 events). 

\begin{definition}\label{def:unicool}
\emph{Uniform Cooling History}: The grass heating system was inactive the full 6 hours before $t_{ON}$.
\end{definition}

\begin{definition}\label{def:intercool}
\emph{Intermittent Cooling History}: The grass heating system was intermittently active for at most 3 hours during the 6 hours before $t_{ON}$.
\end{definition}

\subsubsection{Energy Consumption}
\begin{figure}[t]
\centering
\begin{tabular}{ccc}
\subfigure[In relation to $T_{Gset}$.\label{fig:demand_events_supply}]{\includegraphics[width=0.3\textwidth]{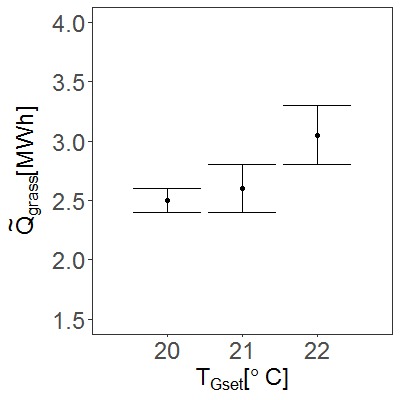}} &
\subfigure[In relation to $\Delta~T_{root,ext}$.\label{fig:demand_events_heatloss}]{\includegraphics[width=0.3\textwidth]{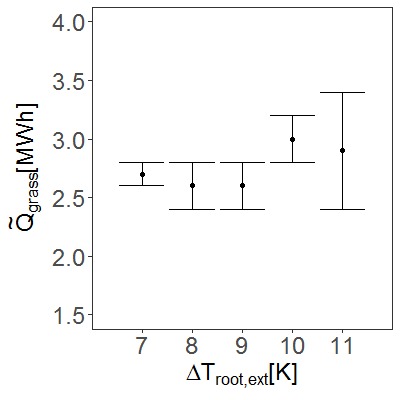}} &
\subfigure[In relation to $\Delta~T_{Gset,root}$.\label{fig:demand_events_XSheatloss}]{\includegraphics[width=0.3\textwidth]{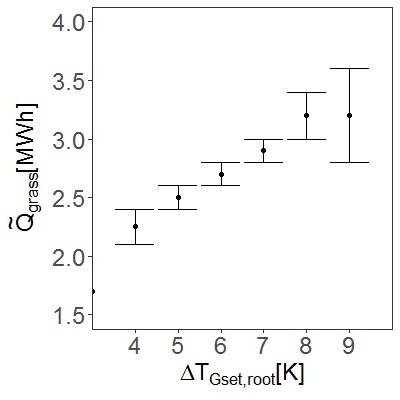}} 
\end{tabular}
\caption{Confidence intervals of median energy consumption for 6 hour heating events using Definition \ref{def:unicool}.}\label{fig:demand_eventsCI}
\end{figure}

\begin{figure}[t]
\centering
\begin{tabular}{ccc}
\subfigure[In relation to $T_{Gset}$.\label{fig:demand_events_supply}]{\includegraphics[width=0.3\textwidth]{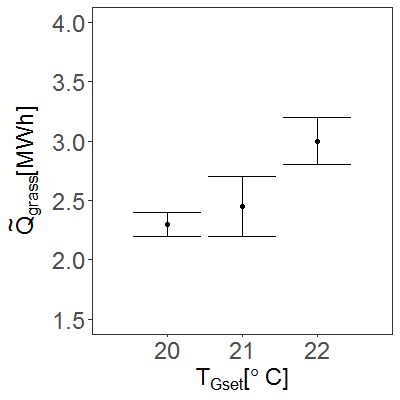}} &
\subfigure[In relation to $\Delta~T_{root,ext}$.\label{fig:demand_events_heatloss}]{\includegraphics[width=0.3\textwidth]{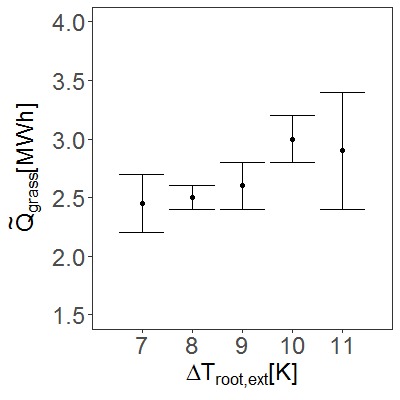}} &
\subfigure[In relation to $\Delta~T_{Gset,root}$.\label{fig:demand_events_XSheatloss}]{\includegraphics[width=0.3\textwidth]{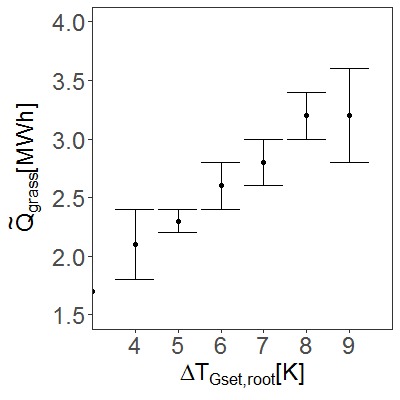}} 
\end{tabular}
\caption{Confidence intervals of median energy consumption for 6 hour heating events using Definition \ref{def:intercool}.}\label{fig:demand_eventsCI_inter}
\end{figure}

Using the heating events of each definition, Fig.~\ref{fig:demand_eventsCI} and \ref{fig:demand_eventsCI_inter} depict the intervals of 95\% confidence for the estimate of the true median energy consumption ($\tilde{Q}_{grass}$) per heating activation event derived by bootstrapping. Overall, Definition \ref{def:unicool} yields slightly narrower confidence intervals than Definition \ref{def:intercool} as the corresponding data set is more uniform. 
The figures exhibit a clear increase of energy for higher $T_{Gset}$ and higher $\Delta~T_{Gset,root}$. For both event history definitions, there is significant overlap of the confidence intervals. Thus, while the intervals are suitably narrow for heating control strategies to take informed heating operation decisions, numerical optimization cannot be applied to select $T_{Gset}$. The underlying trend of the confidence intervals concerning $\Delta~T_{root,ext}$ is less pronounced. The application of regression models (MLP, DBN) to heating event energy data did not produce satisfactory results. The corresponding analysis is omitted for brevity.

%\emph{[For the considered 6 hour durations of heating events, the impact of $T_{external}$ on energy consumption per heating event is less clear than that of $T_{root}$.]} 

\subsubsection{Thermal Effects}\label{sec:heating_thermal_effect}
Statistical inference for the evolution of $T_{root}$ (denoted $\Delta~T_{root}$) related to heating system activation at time $t_{ON}$ does neither produce sufficiently narrow confidence intervals for Definition \ref{def:unicool} nor for Definition \ref{def:intercool} to formulate control strategies. This section omits the associated analysis for brevity and focuses on MLP and DBN trained with the available heating activation events for both definitions. Under the assumption that weather forecast information helps regression accuracy, the effect of a perfect forecast of $T_{external}$ is studied for the MLP and DBN models. That leads to a total of 8 different combinations of the conceptual choices. For each concept ($\{\text{MLP}, \text{DBN}\}\times\{\text{uniform}, \text{intermittent}\}\times\{\text{perfect forecast}, \text{no forecast}\}$), the grid search results in several regression models with similar performance. Fig.~\ref{fig:heating} presents results of an MLP of two hidden layers with 50 nodes each, and of a DBN with two hidden layers of 100 and 50 nodes, respectively. The MLP learning rate is 0.1. The DBN pre-training learning rate is 0.001, the learning rate 0.1. Both models rely on the 36 preceding $T_{root}$ and $T_{external}$ measurements to predict the next 36 $\Delta~T_{root}$ from $t_{ON}$. 
The models taking weather forecast information into account use the next 6 hours of $T_{external}$ (downloaded from \cite{wunderground}, interpolated to 10-minute intervals) as additional input features. 
The figure illustrates the regression accuracies for uniform and intermittent cooling histories and whether or not using perfect forecast information. From Fig.~\ref{fig:heating} follows:
\begin{itemize}
	\item the DBN consistently outperforms the MLP over H, but by less than $0.1K$;
	\item the lowest RMSE for a 6-hour heating point estimate ($h=H=36$) is $0.4K$ using uniform cooling history data;
	\item the delayed impact of $T_{external}$ on $T_{root}$ limits the effect of using accurate air temperature forecasts on the RMSE, hence the small accuracy improvements;	
	\item even for small $h$ all $RMSE>0K$, which we attribute to (a) the fact that potentially interesting soil parameters such as humidity are not measured and (b) the temperature sensor's measurement resolution;
	\item using uniform cooling history improves regression accuracy as the data is more uniform despite reducing the data sets' sizes.	
\end{itemize}

\begin{narrowfig}{0.5\textwidth}[t]
\centering
\includegraphics[width=0.5\textwidth]{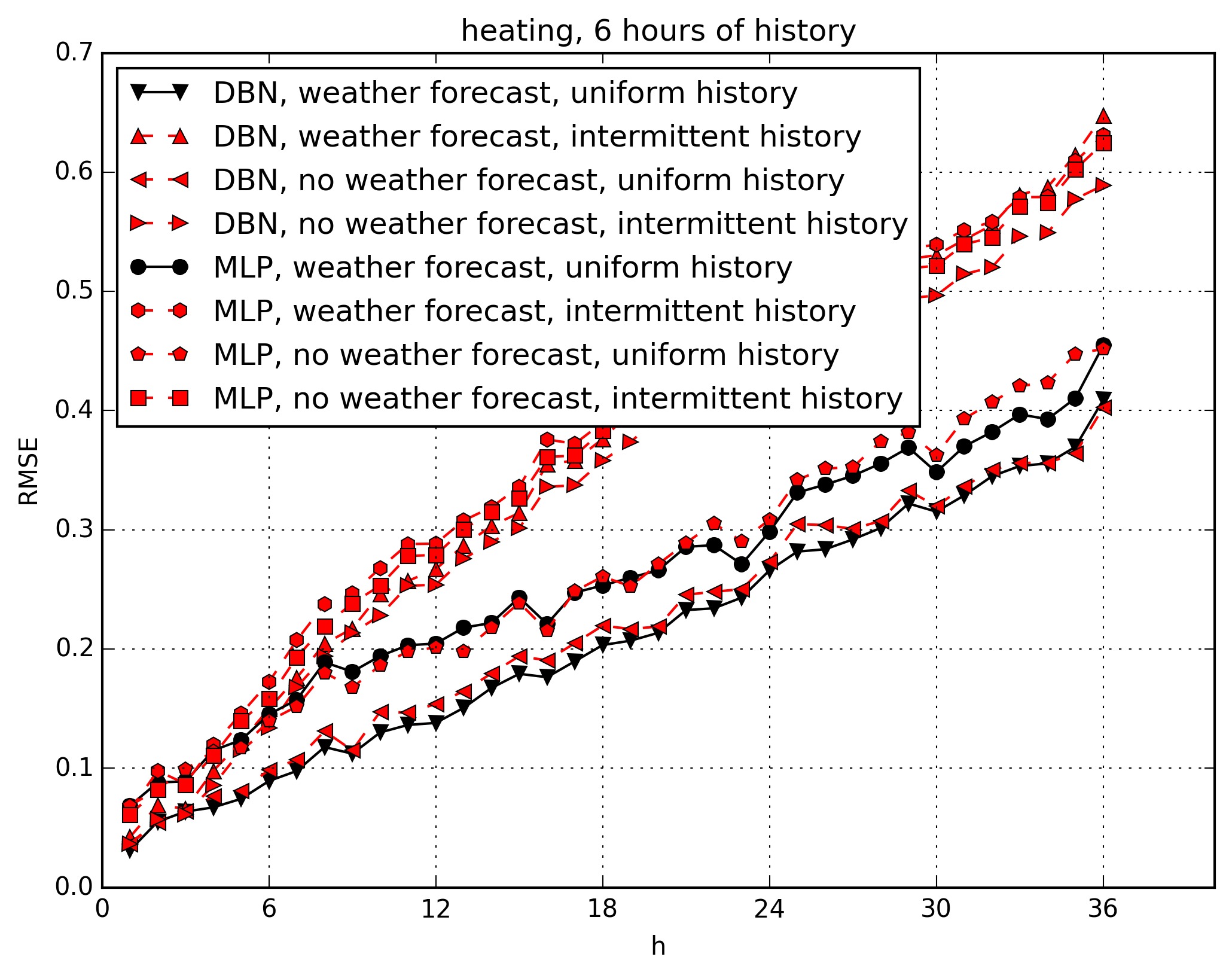}
\caption{$RMSE^h$ for up to 6 hours ($H$=36) of heating for MLP (circle, hexagon, pentagon, square) and DBN (triangles), with and without using perfect weather forecast, in relation to cooling history definition. Black solid lines represent the best performing MLP (circle) and DBN (triangle down): using weather forecast and uniform cooling history. The red dashed lines represent the MLPs and DBNs achieving lower accuracies.}\label{fig:heating}
\end{narrowfig}

As the DBN exhibits a slightly lower and less variable RMSE than the MLP, the predictive heating strategies use a DBN with 6 hours of uniform cooling history, taking into account weather forecast information. Relative to the temperature target band's width of 2K, the resulting RMSE is acceptable - especially the first 3 hours of predictions ($h\leq 18$) exhibit a relative RMSE $\lessapprox$ 10\% compared to the target band's width. Moreover, it is feasible to execute the regression models each time step (10 minutes), which mitigates possible prediction errors of earlier time steps. %Also, the control strategies can take the regressed $\Delta~T_{root}$ trajectory into account to increase the stability of control decisions. 

\subsection{The Effects of Heating Deactivation}\label{sec:heating_deactivation}
\subsubsection{Statistical Considerations, Defining Cooling Events}
Similar to Section \ref{sec:heating_activation}, the intention of this subsection is to build an understanding of the soil temperature dynamics when heating deactivates at $t_{OFF}$. The following characteristics establish independence among data samples of heating deactivation events.

\begin{itemize}
\item $t_{OFF}$ marks the start of a period with inactive grass heating of 6 hours or longer. 
\item A period of active heating precedes $t_{OFF}$ to mitigate any previous cool-down period. That allows considering the events as mutually independent.
\end{itemize}

Considering the second aspect, this subsection studies two alternate ways of defining the period preceding $t_{OFF}$, both ensuring event independence. Definition \ref{def:uniheat} selects 72 heating system deactivation events (the regression model training set contains 43 events). Definition \ref{def:interheat} selects 99 events (leading to a training set size of 60). 

\begin{definition} \label{def:uniheat}
\emph{Uniform Heating History}: The grass heating system was active the full 6 hours before $t_{OFF}$.
\end{definition}

\begin{definition} \label{def:interheat}
\emph{Intermittent Heating History}: The grass heating system was intermittently inactive for at most 3 hours during the 6 hours before $t_{OFF}$.
\end{definition}

\subsubsection{Thermal effects}

\begin{narrowfig}{0.5\textwidth}[t]
\centering
\includegraphics[width=0.5\textwidth]{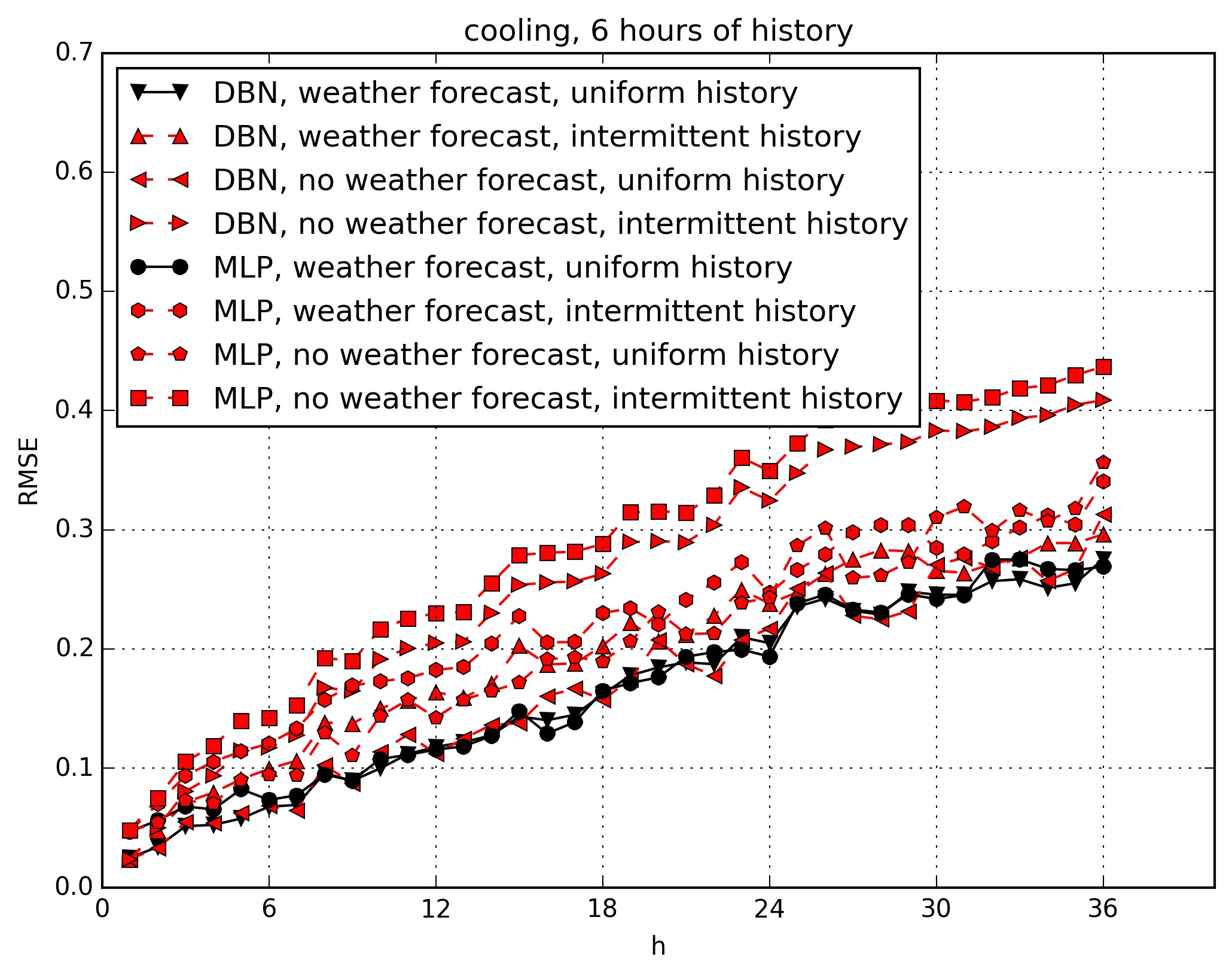}
\caption{$RMSE^h$  for up to 6 hours ($H$=36) of cooling for MLP (circle, hexagon, pentagon, square) and DBN (triangles), with and without using perfect weather forecast, in relation to heating history definition. Black solid lines represent the best performing MLP (circle) and DBN (triangle down): using weather forecast and uniform heating history. The red dashed lines represent the MLPs and DBNs achieving lower accuracies.}\label{fig:cooling}
\end{narrowfig}

Statistical inference for $\Delta~T_{root}$ in relation to $t_{OFF}$ does neither produce sufficiently narrow confidence intervals for Definition \ref{def:uniheat} nor for Definition \ref{def:interheat} to formulate control strategies. The associated analysis is omitted for brevity. 
In analogy to the scenario of heating activation, this subsection describes the performance of MLP and DBN models for predicting $\Delta~T_{root}$ under the assumption of a heating switch-off event.  %Due to this non-linearity the MMLR approach does not produce satisfactory results and we omit these for brevity.  
The parameter grid search returned several models of similar performance. Fig.~\ref{fig:cooling} presents the results of an MLP consisting of two hidden layers with 60 and 50 nodes, and a DBN with two hidden layers of 90 and 55 nodes. The MLP learning rate is 0.13. The DBN’s pre-training learning rate is 0.001, its learning rate 0.08. Both models rely on the 36 preceding $T_{root}$ and $T_{external}$ measurements to predict the next 36 $\Delta~T_{root}$ from $t_{ON}$. Analogous to Section \ref{sec:heating_thermal_effect}, the models relying on weather forecast information additionally use 6 hours of $T_{external}$ prediction. The figure presents the accuracies for predicting 6 hours of continuous cooling using both Definition \ref{def:uniheat} and Definition \ref{def:interheat} for heating history data, with and without taking perfect $T_{external}$ forecasts into account. It shows that MLP and DBN models are comparable in accuracy. Moreover, the figure illustrates that 
\begin{itemize}	
	\item accurate air temperature forecasts tend to reduce the RMSE, in particular when using intermittent heating history data; 
	\item using uniform heating history data leads to the best cooling predictions for \\$h=H=36$ ($RMSE\approx~0.28K$);
	\item similar to Section \ref{sec:heating_activation}, even for small $h$ all $RMSE>0K$ for both model types;
	\item compared to heating predictions (Fig.~\ref{fig:heating}), cooling predictions exhibit a lower RMSE.
\end{itemize}

Considering the temperature target band's width of 2K, the resulting RMSE is sufficiently small to predictively take heating system deactivation decisions - especially the first 3 hours of predictions ($h\leq 18$) exhibit a relative RMSE below 10\% compared to the target band's width. %Moreover, the control strategies can take the regressed $\Delta~T_{root}$ trajectory into account to increase the stability of control decisions. 
As the best MLP and DBN models are of similar performance, the predictive strategies use the latter. The DBN relies on uniform heating history data and takes weather forecast information into account.  

\section{Methodology Step \ref{approach:step5}: Heating Strategies}
\label{sec:intelligence}
This section summarizes the different control strategies developed in \cite{Schmidt:2015:EEG:2821650.2821661} for reference. As mentioned in Section \ref{sec:heating_activation} neither the statistical inference models nor the neural networks achieved satisfactory accuracy for heating event energy predictions. Therefore, we avoid using computational optimization methods in the heating strategies. Every 10 minutes, the CPS control strategies have access to the past and current BMS variables of Table \ref{tab:variables} and to the forecast of $T_{external}$ provided by \cite{wunderground}. The strategies control whether the grass heating system should be active or inactive, as well as the value of $T_{Gset}$. %The strategies' performances are evaluated based on data on the grass heating system's median daily weather-normalized energy consumption ($\tilde{Q}_{grass,HDD7}$) and the fraction of time that $T_{root}$ violates the minimum temperature target (UPR).  

In coordination with staff, based on the ranges of glycol supply temperatures observed in the reference winter, the strategies are allowed to choose $T_{Gset} \leq 22^\circ C$. The lower $T_{root}$ target temperature defines the minimum $T_{Gset}$ permissible. This approach is implemented by a failsafe mechanism in each control strategy after e.g.~executing the predictive regression models of Section \ref{sec:cps_rep}. That software failsafe ensures that regression model errors do not cause excessively high or low $T_{Gset}$ choices. The grass heating system's failsafe described in Section \ref{sec:setting} provides an additional level of protection against overheating and consequently damaging the grass.

\subsection{Basic Control Strategies}
\subsubsection{Basic Strategy B1: Static Supply, On/Off}
This simplest of strategies uses a fixed $T_{Gset}$ during the nighttime. It activates heating when $T_{root}<12^\circ C$ and deactivates heating when $T_{root}>14^\circ C$.

\subsubsection{Basic Strategy B2: Variable Supply}
Strategy \emph{B2} continuously heats during the nighttime with varying $T_{Gset}$. This way, it can gradually react to the system's environmental context - the weather impact on $T_{root}$. The strategy relies on the empiric findings of thermal behavior \cite{Schmidt:2015:EEP:2735960.2735978} to vary $T_{Gset}\in[12^\circ C, 22^\circ C]$. Specifically, steps 3 and 4 modify the steepness of the heating and cooling curves to avoid violating the target band due to the heating system's (and the soil's) thermal inertia:

\begin{enumerate}
\item Set $T_{Gset}=22^\circ C$, if $T_{root}<12^\circ C$
\item Set $T_{Gset}=12^\circ C$, if $T_{root}>14^\circ C$
\item Increase $T_{Gset}$ by $0.5K$, if $T_{root}<12.5^\circ C \wedge T_{Gset}<22^\circ C$  
\item Decrease $T_{Gset}$ by $0.5K$, if $T_{root}>13.5^\circ C \wedge T_{Gset}>12^\circ C$ 
\end{enumerate}

\subsubsection{Basic Strategy B3: Pre-Heating}
This strategy focuses on countering day cool-down due to the operational condition that the heating is inactive during the daytime. The intention is to ensure that $T_{root}$ is near the upper limit of the target band at the end of each nightly heating phase. The strategies \emph{B1} and \emph{B2} do not explicitly account for this. In essence, \emph{B3} mimics the current best practice of nighttime pre-heating, but with a much faster reaction time than the human control during the reference period. The strategy reuses \emph{B2} during the first hours of heating. Taking into account the cooling speeds observed, \emph{B3} modifies the third and fourth steps of \emph{B2} between 04:00 and 06:00 based on empirical observations about the heating system's inertia, resulting in the desired pre-heating at the end of each night: $T_{root} \in [13.5^\circ C, 14^\circ C]$. 

\begin{enumerate}
%\item Set $T_{Gset}=22^\circ C$, if $T_{root}<12^\circ C$
%\item Set $T_{Gset}=12^\circ C$, if $T_{root}>14^\circ C$
\item[(3)] Increase $T_{Gset}$ by $2.0K$, if $T_{root}<13.50^\circ C \wedge T_{Gset}~\leq~20^\circ C$ 
\item[(4)] Decrease $T_{Gset}$ by $2.0K$, if $T_{root}>13.75^\circ C \wedge T_{Gset}~\geq~14^\circ C$ 
\end{enumerate}

\subsection{Advanced Control Strategies: Operational Context and Predictive Control}

\subsubsection{Strategy D: Introducing Daytime Heating}
Daytime heating is delicate as the heating capacity constraints are known to negatively affect some of the attached thermal sub-systems such as the office heating. 
That happens when the output of the stadium’s gas boilers is insufficient for the overall heating demand, causing the main supply circuit's temperature $T_{supply}$ to drop. As a consequence, the Commerzbank Arena's physical heating distribution system prioritizes the grass heating system over other heating systems. Therefore, the control strategy must take into account the grass heating system’s \emph{operational context}, i.e.~the operational situation of other systems to avoid thermal supply scarcity when changing the grass heating paradigm to include the daytime. In this specific setting, $T_{supply}$ is a good indicator for situations of thermal peak demand, and thus, the grass heating system's operational context is sufficiently well captured by that single variable. Discussions with staff defined a threshold value of $T_{supply}=80^\circ C$, which ensures the standard operation of the other heating systems. When $T_{supply}$ undercuts the threshold during the daytime, the strategy deactivates the grass heating immediately and reduces $T_{Gset}$ to the minimum. After this kind of deactivation, when $T_{supply}\geq~80^\circ C$, $T_{Gset}$ is slowly increased again. This slow ramp up mechanism in response to peak demand prevents oscillations of grass heating system operation that would unnecessarily stress the system. In other words, the control strategy addresses a limitation of the physical heating distribution system in case $T_{root}$ is sufficiently high: it reacts to scarcity and limits the grass heating consumption accordingly. 
As the heating demand of offices and other arena areas peaks at office hour start, the strategy reuses \emph{B3} during the nighttime. That ends each night with a pre-heat cycle resulting in higher $T_{root}$, enabling a safe deactivation of the grass heating at office day start without risking under-performance. For daytime operation when $T_{supply} > 80^\circ C$, the variable supply logic \emph{B2} is reused.% - as also illustrated in Fig.~\ref{fig:d1Algorithm}.

\subsubsection{Strategy Dmod: Modified Daytime Heating}
After introducing the paradigm change to daytime heating in \emph{D}, this subsection describes a more aggressive strategy \emph{Dmod}. % is depicted in Fig.~\ref{fig:d1modAlgorithm}.  
Its aim is to increase the amount of heating energy used during the day even further to keep $T_{root}$ in the middle of the target band. That flattens the temperature curves and relieves nighttime operation from mitigating cool-down effects, which reduces overall energy consumed. 
During daytime, it differs from \emph{D} twofold: 

\begin{enumerate}
\item[(a)]	\emph{Dmod} applies a lower $T_{supply}$ threshold of $75^{o}C$. 
\item[(b)]	If $T_{external} \geq 5^\circ C$, it uses an even lower $T_{supply}$ back-off threshold of $70^\circ C$.
\end{enumerate}

\subsubsection{Strategy PA1: Predictive Pre-heating}
This strategy is intended to study the effect of using weather forecast information in addition to sensor readings. The heating approach is reverted to the nighttime heating paradigm to isolate the energy gains due to improved forecast accuracy. 
During nighttime, \emph{PA1} uses the trained DBN cooling model of Section \ref{sec:heating_deactivation} to predict a first 6-hour $\Delta~T_{root}$ trend. The combination of this first $\Delta~T_{root}$ forecast with another subsequent 6-hour $\Delta~T_{root}$ forecast leads - due to the saturation effect of cooling curves - to a pessimistic 12-hour $\Delta~T_{root}$ prediction. If this prediction indicates $T_{root}$ undercutting the required minimum temperature, \emph{PA1} uses the heating DBN of Section \ref{sec:heating_activation} to predict $T_{root}$ for 6 hours in different $T_{Gset}$ heating scenarios. \emph{PA1} selects the heating scenario resulting in the highest $T_{root}$ without violating the maximum temperature during the next 6 hours. If the remaining time of the nightly heating phase is shorter, \emph{PA1} considers only the first $h$ $T_{root}$ predictions, where $t_{now} + 10 \text{min} \times h \leq \text{06:00}$.
This strategy implies a pre-heating effect similar to \emph{B3}: as the prediction horizon shortens towards the end of the night, \emph{PA1} checks less of the predicted $T_{root}$ for violating the target band's upper threshold. Therefore, higher $T_{Gset}$ are selected automatically. In winter 2015/2016, another experiment with a $T_{root}$ target band lowered by 2K was executed (denoted \emph{PA1\textsuperscript{*}}). 
  
\subsubsection{Strategy PA2: Predictive Pre-heating with Longer Forecast Horizon}
Based on \emph{PA1}, the slightly varied strategy \emph{PA2} directly produces a 12-hour $T_{root}$ cooling trend with a single trained DBN ($RMSE=0.65\pm0.2 K$, not presented in Section \ref{sec:cps_rep} for brevity) instead of forecasting two consecutive 6-hour horizons. 

\section{Methodology Step \ref{approach:step6}: Experimental Validation and Discussion}
\label{sec:discussion}
\subsection{Experiments' Impacts}
\label{sec:impact}
Reference winter 2013/2014 accumulated $\text{HDD7}=327~\text{DD}$, while the 5-year average $\text{HDD7}=574~\text{DD}$ \cite{wunderground}. Fig.~\ref{fig:winter1314_temps} shows the reference period's $T_{external}$ and $T_{root}$. During this time, the Commerzbank Arena's grass heating system consumed 795 MWh, representing 20\% of the stadium's overall gas consumption. That is equivalent to 152 t CO$_{2}$ emissions \cite{energiedatengesamt}.
Fig.~\ref{fig:winter1314_grass_root_temps} shows that 85.9\% of the time $T_{root}$ violated the target band by exceeding $14^\circ C$.  According to Fig.~\ref{fig:refHistogram} there was much wasteful heating operation presenting savings potential. The wasteful operation stems from the grass heating control system being mostly driven by $T_{external}$ instead of $T_{root}$, by inconsistent control strategy changes, and by the staff's preference for conservative operation settings as it cannot continuously monitor and adapt the heating system operation. However, throughout the reference winter, the minimum temperature was not violated, i.e.~UPR=0\%. %: $T_{root}>14^\circ C$ for 53\% of the system's total runtime. 

\begin{table}[t]
\small
\centering
\tbl{Confidence Intervals (95\% level) for median daily grass heating energy consumption and with median normalized daily grass heating energy consumption for $\text{HDD7}>0$. As the Shapiro-Wilk test rejected the normality hypothesis for most of the data series, bootstrapping was applied to all cases for consistency. For experiment \emph{Dmod}, only the February 2015 data is used for confidence interval calculations due to warm weather. This warm weather also prevented \emph{PA1} and \emph{PA2} to be interpreted from an energy perspective.
\label{tab:performances}}{
\begin{tabular}{cccccc} 
Data Set 	&  $\tilde{Q}_{grass}$   	& $\tilde{Q}_{grass,HDD7}$				& \# Experiment  	& UPR 	& $\overline{\text{HDD7}}\pm~s$  \\ 
					& [MWh] 									& [MWh/DD] 																							& Time Frame					&				& [DD]\\ \hline				
Reference Period  & [5.70,6.60] 		& [2.12,2.88] 																	& 2013/11/01-2014/02/28	& 0\% & $2.17\pm1.92$ \\ 
B1 								& [2.60,4.60] 		& [0.83,1.48]																		& 2014/11/24-2014/12/04	& 11.1\% & $3.34\pm1.46$ \\
B2 								& [3.00,5.20] 		& [0.91,1.11] 																		& 2014/12/04-2014/12/11 	& 12.5\% & $4.37\pm1.02$ \\
B3 								& [4.60,6.20] 		& [1.17,3.01] 																		& 2014/12/11-2015/01/16 & 7.1\% & $3.67\pm2.86$\\
D 								& [5.80,6.80] 		& [1.02,1.41] 														 				& 2015/01/16-2015/02/11 & 2.3\% & $5.55\pm1.81$\\
Dmod\textsuperscript{*}& 	[4.80,6.10] & [1.00,1.29]														& 2015/02/11-2015/03/11 & 0\% & $4.13\pm1.45$\\ 
PA1 							& NA & NA  														 											& 2015/03/11-2015/03/18 	& 5.4\% & $2.00\pm0.98$\\
PA2 							& NA & NA 																 												& 2015/03/18-2015/03/31	& 0.3\% & $1.23\pm1.19$\\ 
(B1,B2,B3,D,Dmod\textsuperscript{*}) & [4.60,5.60] & [1.06,1.33] 	 									& 2014/11/24-2015/03/11	& 5.2\% & $4.07\pm2.40$ \\ 
(D,Dmod\textsuperscript{*}) &	[4.80,6.10] & [1.00,1.28]  							&  2015/01/16-2015/02/28 & 1.5\% & $5.02\pm1.81$\\
PA1\textsuperscript{*} & [1.09,1.40] & [0.57,0.77]   								&  2015/12/21-2016/01/12 	& 4.5\% & $2.03\pm1.63$\\
\hline 
\end{tabular}}
\normalsize
\end{table}

\begin{figure}[t]
\centering
\begin{tabular}{cc}
\subfigure[Winter 2014/2015.\label{fig:airtemps1415}]{
\includegraphics[width=0.45\textwidth]{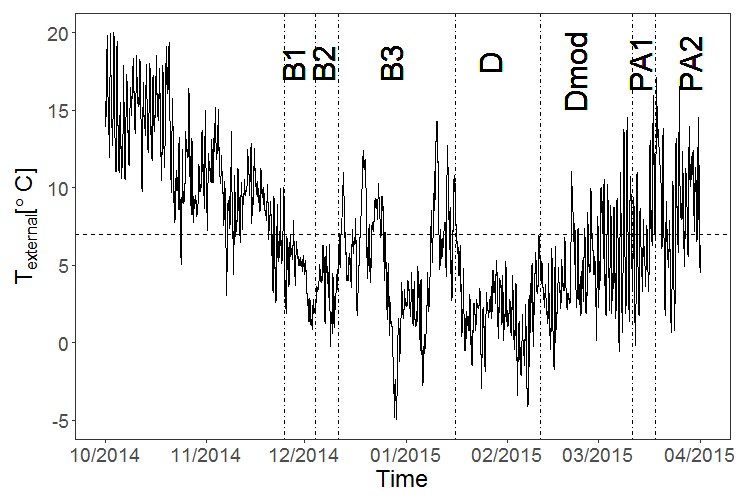}} &

\subfigure[Winter 2015/2016.\label{fig:airtemps1516}]{
\includegraphics[width=0.45\textwidth]{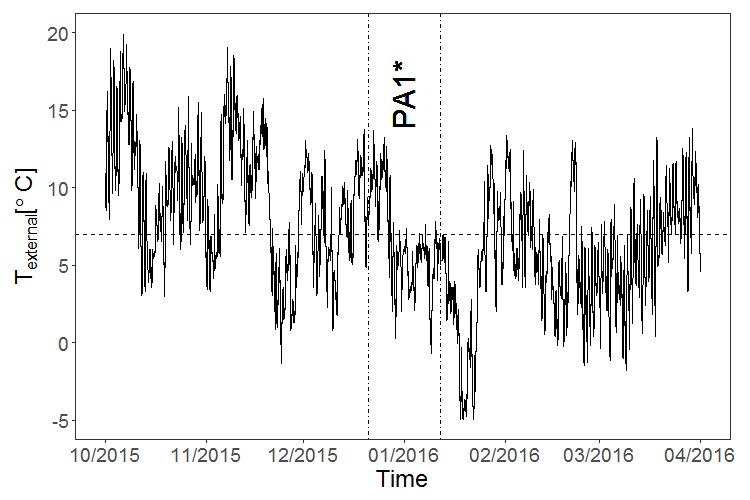}}
\end{tabular}
\caption{Recorded $T_{external}$ of experiment winters. The dashed horizontal lines indicate the grass heating switch-on threshold temperature of $7^\circ C$.}\label{fig:airtemps_experiments}
\end{figure}

\begin{figure}[t]
\centering
\begin{tabular}{cc}
\subfigure[Winter 2014/2015 $T_{root}$ evolution, $T_{Gset}$ decided algorithmically.\label{fig:winter1415_grass_root_temps}]{
\includegraphics[width=0.45\textwidth]{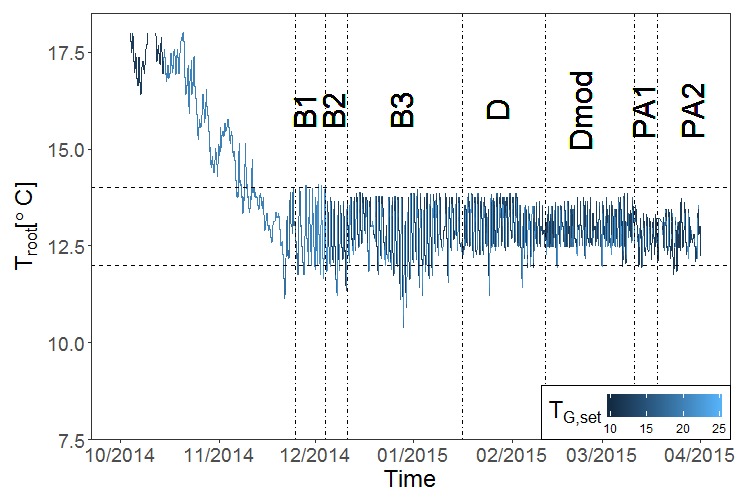}} &

\subfigure[Winter 2015/2016 $T_{root}$ evolution, $T_{Gset}$ decided algorithmically.\label{fig:winter1516_grass_root_temps}]{
\includegraphics[width=0.45\textwidth]{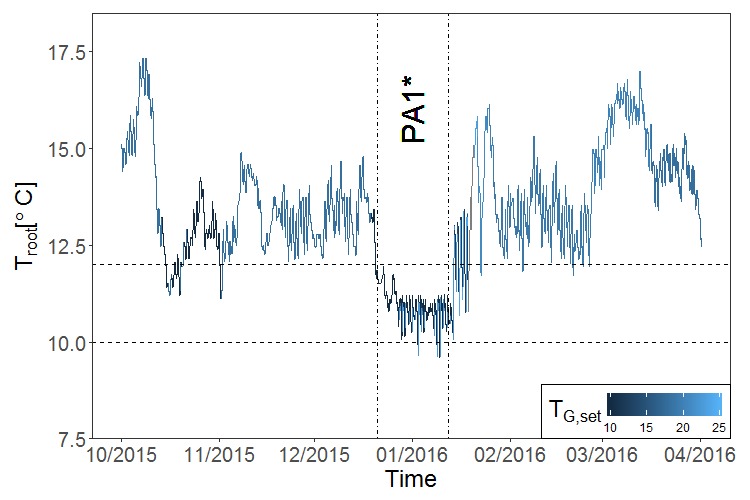}} 
\end{tabular}
\caption{Grass root temperatures during experiments. The dashed horizontal lines indicate the temperature target bands.\label{fig:exp_temps}}
\end{figure}

Fig.~\ref{fig:airtemps_experiments} and \ref{fig:exp_temps} depict the measured $T_{external}$ and $T_{root}$ during the winters 2014/2015 and 2015/2016.  Further, they indicate each experiment's execution period. While the heating season officially lasts from October to March, Fig.~\ref{fig:airtemps1415} shows that experiments of winter 2014/2015 could only start towards the end of November 2014 due to warm weather. In Winter 2014/2015, pre-heating (\emph{B3}) and daytime heating (\emph{D}, \emph{Dmod}) were prioritized and received more experimentation time as indicated in Fig.~\ref{fig:airtemps1415} and \ref{fig:winter1415_grass_root_temps}. With $\text{HDD7}=479~\text{DD}$ the experimental period 2014/2015 was colder than the reference period but still warmer than the average Frankfurt winter. Fig.~\ref{fig:airtemps1516} and \ref{fig:winter1516_grass_root_temps} show the \emph{PA1} re-execution between 2015-12-21 and 2016-01-12 to confirm the  UPR results of using predictive heating and also to quantify the effect of lowering $T_{root}$ on energy consumption. For this experiment, the $T_{root}$ target band was lowered by 2K to $10^\circ C\leq~T_{root}\leq~12^\circ C$.  %Fig.~\ref{fig:airtemps1516} shows the measured winter 2015/2016 weather conditions and indicates the \emph{PA1\textsuperscript{*}} execution period. 

Table \ref{tab:performances} summarizes the collected data. It details energy consumption confidence intervals inferred for a single median for reference winter 2013/2014, the experiments of winter 2014/2015, and for \emph{PA1\textsuperscript{*}}. Experiments \emph{Dmod}, \emph{PA1}, and \emph{PA2} suffered from warm weather in February and March 2015. That precludes an interpretation of the energy consumption of experiments \emph{PA1} and \emph{PA2}, as well as the second half of experiment \emph{Dmod}. Thus, the table presents \emph{B1}, \emph{B2}, \emph{B3}, \emph{D}, and the February sub-period of \emph{Dmod} (indicated by \textsuperscript{*}) as the seasonal aggregate statistics (denoted (\emph{B1,B2,B3,D,Dmod\textsuperscript{*}})) of winter 2014/2015. The table also singles out the effect of the heating paradigm change towards daytime heating by aggregating \emph{D} and \emph{Dmod\textsuperscript{*}}. Table \ref{tab:performances} also provides the UPR observed, the HDD normalization factor (mean and standard deviation) from \cite{wunderground}, and the different experiment execution periods. As days with switching experiments do not affect UPR statistics, these are calculated by taking the accurate experiment activation and deactivation times. As the energy considerations rely on the median daily consumption, days with experiment switch-over are excluded from the confidence interval calculations as well as from the statistics of HDD7. 

Fig.~\ref{fig:exp_temps} confirms the low UPR values in Table \ref{tab:performances} for the experiments: most of the time, the supervisory control strategies met the grass root temperature bands. Fig.~\ref{fig:grass_root_boxplots} supports this by presenting $T_{root}$ during the reference period and the two experimental winters. The boxplots show that the supervisory control experiments (second and third groups) kept $T_{root}$ in a much tighter band than the staff-controlled heating operation (leftmost and rightmost groups). For the experiments, $T_{root}$ is predominantly inside the respective target band. Both the aggregate of winter 2014/2015 experiments (\emph{B1,B2,B3,D,Dmod\textsuperscript{*}}), as well as the winter 2015/2016 experiment \emph{PA1\textsuperscript{*}} achieve a low UPR of approximately 5\%. The subset of experiments implementing the paradigm change to daytime heating, enabled by being aware of the grass heating system's operational context, reported even lower UPR: the strategies combine nightly pre-heating with the ability to draw on unused thermal supply capacity during the day. 

\begin{narrowfig}{0.55\textwidth}
\centering
\includegraphics[width=0.55\textwidth]{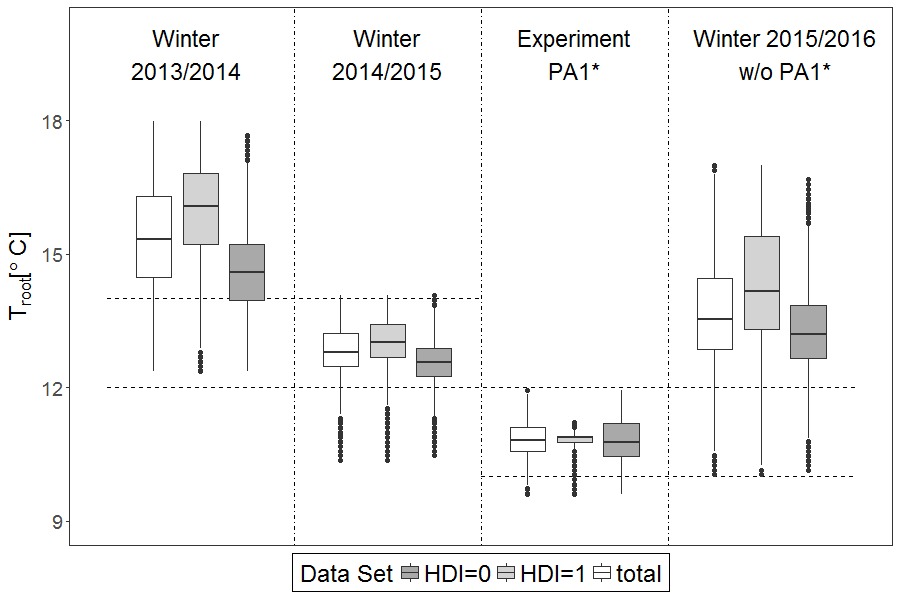}
\caption{Boxplots of $T_{root}$ regardless of HDI status flag ("total"), for active grass heating ("HDI=1") and for inactive grass heating ("HDI=0") of reference period (first group), experimental winter 2014/2015 (second group), of experiment \emph{PA1\textsuperscript*} (third group), and of winter 2015/2016 excluding \emph{PA1\textsuperscript{*}} (fourth group). Dashed horizontal lines indicate the respective $T_{root}$ target ranges.}\label{fig:grass_root_boxplots}
\end{narrowfig}

Experiment \emph{B3} demonstrates that the current best practice of human-controlled nighttime pre-heating is reproducible with higher energy efficiency while keeping UPR below 10\%. Similarly, experiments \emph{D} and \emph{Dmod\textsuperscript{*}} prove that control strategies being aware of a systems operational context can have extremely positive effects. They save energy and exhibit a more consistent consumption (i.e.~fewer outliers), expressed by narrower confidence intervals of normalized median energy than the reference pre-heating strategy. Additionally, this context awareness enables a heating paradigm change while avoiding well-known operational problems of heating bottlenecks negatively affecting other arena heating systems.

While Fig.~\ref{fig:exp_temps}, Fig.~\ref{fig:grass_root_boxplots}, and Table \ref{tab:performances} show that the experiments mostly met the grass root temperature bands, it is important to understand the reasons for its violation:
\begin{itemize}
\item Strategies B1 and B2 do not implement pre-heating. Thus, daytime cool-down increased these experiments' UPRs. In particular, \emph{B1} produced too steep heating curves for higher $T_{Gset}$ at the beginning of mild nights. That resulted in early heating deactivation during the nights so that at the respective night-ends $T_{root}$ was too low for the soil's thermal inertia to keep $T_{root}\geq~12^\circ C$ over daytime.
\item On request of operational staff, experiments were partially interrupted on match days: 2014-12-07, 2015-01-24, and 2015-02-03. That could be addressed reflecting the match schedule in the pre-heating strategy. For example, \emph{D} achieved UPR=0\% when excluding the days with a delayed start from the analysis. 
\item On freezing days, e.g.~during the period of several days around 2014-12-29, the nightly pre-heating towards $14^o$C did not suffice. This situation could be remedied by switching to daytime heating, or by increasing the permissible target band's upper limit depending on weather forecast information.
\item In some cases, the temperature predictions of \emph{PA1\textsuperscript{*}} underestimated cool-down effects for the coming day - partly because of inaccurate weather forecasts, but also due to the DBN models' RMSEs. 
\end{itemize}

Feedback from the Commerzbank Arena's staff on the \emph{PA1\textsuperscript{*}} experiment with a 2K reduced temperature target band indicated that due to a very wet winter 2015/2016, grass health worsened despite satisfactory levels of UPR. To reduce the high levels of moisture by evaporation, \emph{PA1\textsuperscript{*}} was stopped and temperature targets were increased in January 2016 (see Fig.~\ref{fig:winter1516_grass_root_temps}). That shows that UPR may not suffice as grass quality indicator in certain environmental conditions. %Also, operational staff needs to be able to adjust the temperature target band as needed for best grass health. 

Statistical inference on \emph{two population medians} allows studying the experiments' effect sizes by quantifying the pairwise differences. Bootstrapping establishes the following findings with 95\% confidence. 
\begin{enumerate}
\item Compared to the winter 2013/2014 period's median daily normalized energy consumption:
\begin{enumerate}
\item The winter 2014/2015 experiments aggregate of \emph{(B1,B2,B3,D,Dmod\textsuperscript{*})} reduced $\tilde{Q}_{grass,HDD7}$ by 1.13-1.35 MWh/DD (39.2-63.7\% compared to the reference period's confidence interval bounds). In an average winter, savings are expected to amount to 648.6-774.9 MWh, confirming earlier findings. The equivalent of CO$_2$ emissions saved is 123.9-148.0 t. %To verify, $\tilde{Q}_{grass,HDD15}$ was reduced by 0.17-0.22 MWh/DD (26.2-40.7\%).

\item The nighttime pre-heating strategy inspired by the status quo operation (\emph{B3}) reduced $\tilde{Q}_{grass,HDD7}$ by 0.44-0.83 MWh/DD (15.3-39.2\%). That highlights the positive impacts of algorithmic control mimicking best practices. In an average winter, savings of 252.6-476.4 MWh (48.2-91.0 t CO$_2$) can be expected.  %To verify, $\tilde{Q}_{grass,HDD15}$ was reduced by 0.13-0.18 MWh/DD (20.0-33.3\%).

\item The daytime heating \emph{(D,Dmod\textsuperscript{*})} reduced $\tilde{Q}_{grass,HDD7}$ by 1.19-1.40 MWh/DD (41.3-66.0\%). That demonstrates the power of a paradigm change to daytime heating, enabled by taking system operational context into account. 683.1-803.6 MWh (130.5-153.5 t CO$_2$) of savings can be expected in an average winter. %To verify, $\tilde{Q}_{grass,HDD15}$ was reduced by 0.17-0.22 MWh/DD (20.0-33.3\%).

\item The predictive nighttime pre-heating with lowered $T_{root}$ target band \emph{PA1\textsuperscript{*}} reduced $\tilde{Q}_{grass,HDD7}$ by 1.57-1.80 MWh/DD (54.5-84.9\%). In an average winter, savings are expected to reach 901.2-1,033.2 MWh (172.2-197.3 t CO$_2$). That provides a quantification of the effect of a lowered temperature target band while using only predictive nightly pre-heating. Even stronger effects are anticipated when used in combination with daytime heating. %To verify, $\tilde{Q}_{grass,HDD15}$ was reduced by 0.25-0.30 MWh/DD (38.5-55.6\%).
\end{enumerate}

\item Compared to a simple automated best practice strategy (\emph{B3}), the benefits of a daytime heating strategy aware of the grass heating system's operational context \emph{(D,Dmod\textsuperscript{*})} can also be inferred. The inference shows the change of heating paradigm towards daytime heating reduced $\tilde{Q}_{grass,HDD7}$ by 0.53-0.76 MWh/DD (17.6-65.0\%, 304.2-436.2 MWh, 58.1-83.3 t CO$_2$ in an average winter). These savings are comparable to the savings of \emph{B3} over the current status-quo operation. Additionally, daytime heating improved the UPR.%To verify, $\tilde{Q}_{grass,HDD15}$ was reduced by ... MWh/DD (...\%).

\item Experiment \emph{PA1\textsuperscript{*}} provides an indication of the magnitude of the additional energy savings potential associated with lowering the target temperature band by 2K. \emph{PA1\textsuperscript{*}} reduced daily $\tilde{Q}_{grass,HDD7}$: 
\begin{enumerate}
\item  by 0.37-0.52 MWh/DD, compared to \emph{(B1,B2,B3,D,Dmod\textsuperscript{*})} ;
%\item daily $\tilde{Q}_{grass,HDD15}$ by xxx MWh/DD, compared to \emph{(B1,B2,B3,D,Dmod\textsuperscript{*})} ;
\item by 0.33-0.45 MWh/DD, compared to \emph{(D,Dmod\textsuperscript{*})};
%\item daily $\tilde{Q}_{grass,HDD15}$ by xxx MWh/DD, compared to \emph{(D,Dmod\textsuperscript{*})};
\item by 0.90-1.16 MWh/DD, compared to \emph{B3}.
%\item daily $\tilde{Q}_{grass,HDD15}$ by xxx MWh/DD, compared to \emph{B3}. 
\end{enumerate}

\end{enumerate}

During the daytime heating experiments \emph{D} and \emph{Dmod}, the control strategies deactivated the grass heating 51 and 48 times due to violations of the $T_{supply}$ threshold. Because of this reactivity to peak load conditions no other stadium heating system suffered from heating shortages. Thus, the grass heating system could be served satisfactorily without adversely impacting other systems. In theory, these strategies should already be able to control the grass heating also on days with soccer events satisfactorily. By integrating match plan information e.g.~to result in higher nighttime pre-heating temperatures, also pro-active load shedding should be feasible.

The reported experimental results are of practical and statistical significance. The energy savings have been achieved consistently in two consecutive winters by applying data-driven strategies using a cyber-physical system that integrates existing building automation infrastructure. The control strategies activated neither the software nor the system's failsafe.
The achieved savings are complementary to refurbishment measures. For example, \cite{Smulders} suggests two measures for the Commerzbank Arena's grass heating system to save energy:
\begin{enumerate}
\item serving the heat exchanger by the arena wide thermal return circuit instead of the supply circuit; and
\item dividing the grass heating system into four sub-circuits for higher temperature control resolution of the soccer pitch.
\end{enumerate}
In combination, these could save approximately 8\% of annual thermal energy (381 MWh/year).
While the first measure is fully compatible with the presented CPS approach, the second would require a minor adaptation of the control strategy definitions to  operate all four grass heating system circuits individually.

\subsection{Hypotheses}
The experiments confirmed the formulated hypotheses in Section \ref{sec:problemstatement} as follows:
\begin{enumerate}
\item \emph{The automation of currently manual supervisory control decisions improves efficiency in daily operation as less conservative operational settings are needed.}
	
Statistical inference of two medians shows with 95\% confidence that experiment \emph{B3} reduced energy consumption by 15.3-39.2\%.

\item \emph{Predictive and context-aware control strategies can mitigate heating shortages and further improve the building's operational efficiency.}

The experiments with predictive and context-aware control strategies mitigated negative effects of heating shortages on other heating sub-systems, saved energy, and achieved satisfactory UPR.

\end{enumerate}

\subsection{Limitations}
The grass heating system's energy meter could not be modeled with satisfactory accuracy, preventing the use of optimization techniques in control strategies. It remains for future work to evolve the strategies using optimization.

As a result of prior work, the strategies account for weather through $T_{external}$. Additional studies are needed to identify possible efficiency improvements by reflecting other parameters such as humidity, solar radiation, or wind speed. Also, more investigation is required whether to use additional soil parameters in the algorithmic consideration of grass quality.

As mentioned in Section \ref{sec:requirements}, the German best practice stadium operation does not account for the biologically required $T_{root}$ range. However, comparing the reference period's energy consumption to the experiments may be considered inappropriate because the biggest share of the experiments' savings stems from lowering $T_{root}$. Therefore, the comparison to \emph{B3} provided in Section \ref{sec:impact} provides valuable insights: the daytime heating experiments \emph{(D,Dmod\textsuperscript{*})} show substantial savings of 300-400 MWh in an average winter. That quantifies the additional benefit of moving from an automated best practice control strategy to a control strategy aware of the system's operational context. The associated additional savings are as large as those of moving from manual best practices to automating these.  

 The available data for training, validating, and testing the MLP and DBN is small due to the event definition used in sections \ref{sec:heating_activation} and \ref{sec:heating_deactivation}. Hence, the models' test set performances might improve with access to more data. However, the potential of increasing energy efficiency associated with improved regression performance is considered small, because the achieved RMSE relative to the $T_{root}$ target band's width is below 10\% for the first hours of prediction. Moreover, as the data sampling time (10 minutes) exceeds the regression model execution time (sub-second) by orders of magnitude, inaccurate model predictions and control strategy actions can be rectified in the next time step. Besides, the control strategies could also be evolved to take the predicted $\Delta~T_{root}$ trajectory into account to increase the stability of control decisions. Hence, tuning the model performance is of low priority.

\subsection{Transferability}\label{sec:transfer}
The presented methodology is transferable to other buildings equipped with BMS that can be enabled with standard communication protocols as required. The approach to leverage existing building automation infrastructure and develop CPS control strategies on top is feasible, flexible, and economically as well as ecologically appealing. The preparatory steps to understand the wider system context and the requirements paved the way to identify savings potential and to formulate suitable control strategies. These strategies lead to lower, yet suitable temperature regimes while accounting for operational context to address limitations and shortages intrinsic to the arena's heating system encountered. 
Depending on the building and its BMS, the appropriate protocols may need to be added to the communication platform \cite{cesbpPaperC21}.% to extract data and send actuation commands. 

Typically, heating supply systems are dimensioned based on the \emph{coincidence factor} at design time, i.e.~by estimating the fraction of total sub-system peak demand expected to coincide. When usage patterns change, sub-systems are upgraded, or control strategies are modified, heating shortages may be the unintended consequences. The approach of CPS control strategies \emph{(D,Dmod)} being reactive to operational context information is an effective means to address this.

The German soccer grass mixture is standardized. Hence, the control strategies’ concepts and their target bands are easily transferable. The target temperature range needs adaptation when encountering other mixtures, e.g.~due to different regions’ climatic conditions. 
As the current best practice of stadium operation does not rely on automated control of root temperatures (Section \ref{sec:requirements}), this work's savings is considered as potential savings for other arenas subject to local climatic conditions. 
For the daytime heating strategies, the $T_{supply}$ thresholds depend on the individual thermal distribution system. However, they are straightforward to adjust based on operational experience or system specifications.

Assuming similar grass heating system dimensioning (1.4 MW), a similar thermal exchange between piping and soil, and similar grass root temperature targets as in the Commerzbank Arena, it should be possible to reuse the trained regression models. However, it should be verified that the climatic parameters other than $T_{external}$ have little influence on $T_{root}$.

\section{Conclusion}
\label{sec:conclusion}
Experiments executed in two winter seasons in the Commerzbank Arena in Frankfurt, Germany, yielded results of statistical significance and practical relevance. 
They present a strong case study validating the concept of developing a CPS that integrates pre-existing building instrumentation. That concept enables the realization of a range of different strategies on top of the stadium's BMS to control the major heat-consuming system and assess the associated impacts. The experiments were integrated with daily stadium routine operation and produced high levels of weather-normalized energy savings while maintaining satisfactory grass root temperature levels. In relation to the status quo operation, winter 2014/2015 experiments saved 39.2-63.7\% energy. In average weather conditions, these savings are expected to amount to 650-775 MWh - an equivalent of 124-148 t CO$_2$ emissions. Out of these experiments, daytime heating enabled by the awareness of the heating system's operational context achieved the best results - from the perspective of UPR, energy consumption (savings of 41.3-66.0\%), and from the operational perspective as these experiments mitigated the adverse effects of heating supply shortages. Another branch of strategies focusing on predictive nighttime-only heating control is to our knowledge the first application of Deep Belief Networks to a building's operational data. In this work, these networks outperform standard feed-forward Multi-Layer Perceptrons in heating prediction accuracy and allow formulating predictive grass heating control strategies that operate satisfactorily. Compared to the human controlled status quo operation, the bulk of this work's savings stems from lowered and tightly controlled grass root temperatures. However, when compared to a strategy mimicking human best practice heating, smarter strategies still reduce consumption significantly: in average winter conditions savings of 304.2-436.2 MWh (58-83 t CO$_2$) are expected. Lowering the grass root temperature targets by 2K in winter 2015/2016 increased savings to 54.5-84.9\%. These savings are anticipated to reach 0.9-1.03 GWh (172-197 t CO$_2$) in an average heating season.  

Beyond saving energy, the experiments also provide evidence for the proposed methodology's feasibility. The approach of deploying reactive and predictive control strategies for thermal system operation using a data-driven CPS approach leveraging existing building instrumentation applies to a wide range of other thermal systems and buildings. The analysis of the wider heating system, the application's requirements, and the appropriate definition of a suitable Key Performance Indicator enabled energy savings even in cases where strategies merely mimic the current human operation. The savings stem from the faster reaction times and regular temperature monitoring intervals. When control focuses on individual building systems, aggressive strategies can achieve savings while avoiding adverse effects on other systems - provided that the strategies appropriately take the operational context into account. Taking the Commerzbank Arena as representative soccer stadium, grass heating system operation has a significant savings potential. It is possible to reduce consumption while meeting temperature targets most of the time and keeping the soccer pitch's grass quality at satisfactory levels. In Germany and Austria, professional stadiums are required to use grass heating systems. On an international scale, several stadiums for soccer, rugby, and American football are also known to use grass heating systems. The crucial parts of this work are transferable to these with little or no adaptation.

It is for future work to study the combined effects of daytime heating with lowered target temperatures and predictive nighttime pre-heating. Further, a flexible adaptation of the target band's upper limit should be studied to account for very low-temperature weather forecasts. Another promising direction is to evolve the strategies to using optimization techniques when choosing set-points. That, however, requires further study to improve the heat meter regression accuracies.  
Finally, novel ways to monitor the grass quality directly rather than exclusively focusing on the root temperature should ensure optimal grass quality at lowered temperature levels. 

\small
\bibliographystyle{ACM-Reference-Format-Journals}
\small
\bibliography{acmsmall-sample-bibfile}
                             % Sample .bib file with references that match those in
                             % the 'Specifications Document (V1.5)' as well containing
                             % 'legacy' bibs and bibs with 'alternate codings'.
                             % Gerry Murray - March 2012

% History dates
%\vspace{-0.5cm}
\received{November 2016}{June 2017}{September 2017}

\end{document}